\documentclass{aa}
\usepackage[varg]{txfonts}
\usepackage[T1]{fontenc}
\usepackage{graphicx}	
\usepackage{xcolor}	
\usepackage{amsmath}	
\usepackage{amssymb}	
\usepackage{footmisc}
\usepackage{hyperref}
\hypersetup{colorlinks   = true, citecolor = blue,
            urlcolor=blue,
            linkcolor= blue}
\usepackage{upgreek}
\usepackage{orcidlink}
\usepackage{placeins}
\bibpunct{(}{)}{;}{a}{}{,} 
\usepackage{soul} 
\usepackage{ulem} 
\usepackage{dblfloatfix}
\usepackage{float}
\usepackage{wrapfig} 
\usepackage{multicol}  
\usepackage{textgreek} 	
\usepackage{siunitx}		
\usepackage{booktabs} 	
\usepackage{xargs}                      
\usepackage{epsfig}
\usepackage{color}
\usepackage{multirow}
\usepackage{bigints}
\usepackage{bm,etoolbox}
\usepackage{calligra}
\usepackage{makecell}
\usepackage{comment}

\sethlcolor{cyan}
\defcitealias{deBeer2023}{Paper I}
\defcitealias{Travascio:inprep}{Travascio et al. in prep.}
\defcitealias{Cantalupo:inprep}{Cantalupo et al. in prep.}

\newcommand{\be}{\begin{equation}}
\newcommand{\ee}{\end{equation}}
\def\ba#1\ea{\begin{align}#1\end{align}}

\newcommand{\refeq}[1]{Eq.~(\ref{eq:#1})}          
                  
\newcommand{\refFig}[1]{Figure~\ref{fig:#1}}
\newcommand{\reffig}[1]{figure~\ref{fig:#1}} 
          
\newcommand{\refsec}[1]{section~\ref{sec:#1}}          
 
\newcommand{\refsecs}[2]{sections~\ref{sec:#1}--\ref{sec:#2}}
\newcommand{\refapp}[1]{appendix~\ref{app:#1}}
\newcommand{\refApp}[1]{Appendix~\ref{app:#1}}
\newcommand{\reftab}[1]{table~\ref{table:#1}}

\newcommand{\Om}{\Omega_m}

\newcommand{\OK}{\Omega_K}
\newcommand{\OL}{\Omega_\Lambda}

\newcommand{\lya}{{\rm Ly}\alpha}
\newcommand{\dalya}{{\sc DaLya} }

\def\apjs{APJS}

\def\apjl{APJL}

\def\aap{A\&A}

\begin{document}
\title{Resolving the physics of Quasar Ly$\alpha$ Nebulae (RePhyNe): II. The dense and clumpy CGM of Quasars at \textit{z} $\sim$ 3.5}
\titlerunning{The clumpy CGM of QSOs}

\author{Titouan Lazeyras\orcid{0000-0002-7080-9839}\inst{1}
    \and Sebastiano Cantalupo\orcid{0000-0001-5804-1428}\inst{1}
    \and Andrea Travascio\orcid{0000-0002-8863-888X}\inst{2}
    \and Nicolas Ledos\orcid{0000-0001-9699-8941}\inst{1}
    \and Joop Schaye\orcid{0000-0002-0668-5560}\inst{3}
    \and Gabriele Pezzulli\orcid{0000-0003-0736-7879}\inst{4}
    \and Federico Baraggioni\inst{5,6}
    \and Alessandro Lupi\orcid{0000-0001-6106-7821}\inst{7,8}
    \and Andrea V. Macci\`{o}\orcid{0009-0004-4596-7941}\inst{5,6}
    \and Marta Galbiati\orcid{0000-0002-1843-1699}\inst{2}
    \and Antonio Pensabene\orcid{0000-0001-9815-4953}\inst{9,10}
    \and Giada Quadri\orcid{0009-0005-9819-5620}\inst{1}
    \and Weichen Wang\orcid{0000-0002-9593-8274}\inst{1,11}}
\offprints{T. Lazeyras, \email{titouan.lazeyras@unimib.it}}
\institute{Dipartimento di Fisica ``G. Occhialini'', Universit\`{a} degli Studi di Milano-Bicocca, Piazza della Scienza 3, 20126 Milano, Italy
\and INAF – Osservatorio Astronomico di Trieste, Via G. B. Tiepolo 11, 34143 Trieste, Italy
\and Leiden Observatory, Leiden University, P.O. Box 9513, 2300 RA Leiden, The Netherlands
\and Kapteyn Astronomical Institute, University of Groningen, Landleven 12, 9747 AD Groningen, The Netherlands
\and Center for Astrophysics and Space Science (CASS), New York University Abu Dhabi, PO Box 129188, Abu Dhabi, UAE
\and New York University Abu Dhabi, PO Box 129188, Abu Dhabi, UAE
\and Como Lake Center for Astrophysics, DiSAT, Universit\`{a} degli Studi dell’Insubria, via Valleggio 11, I-22100 Como, Italy
\and INFN, Sezione di Milano-Bicocca, Piazza della Scienza 3, 20126 Milano, Italy
\and Cosmic Dawn Center (DAWN), Copenhagen, Denmark
\and DTU Space, Technical University of Denmark, Elektrovej 327, DK2800 Kgs. Lyngby, Denmark
\and Chinese Academy of Sciences South America Center for Astronomy (CASSACA), National Astronomical Observatories of China, Beijing 100012, China}

\date{Received -- / Accepted --}

\abstract{
The small scale density distribution of the cold circumgalactic medium (CGM) plays a central role in galaxy evolution, yet it is still poorly understood. We provide new constraints on the cold CGM density distribution by comparing the CGM emission from MUSE observations of quasar (QSO) nebulae at $z \sim 3.8$ with mock observations generated from the new {\sc DaLya} simulation suite under the assumption of maximal fluorescence. These are further complemented by mock observations generated from the COLIBRE (m5 and m6 resolutions) and HELLO simulations. We first investigate the distribution of the observed and simulated Ly$\alpha$ surface brightness (SB) values, which directly depend on the cold gas density distribution under our assumption, finding stark differences between observations and upper limits provided by simulations with the exception of the COLIBRE higher-resolution run (m5). The observations generally show a much larger covering fraction of regions with relatively large Ly$\alpha$ SB values and suggest that densities reaching at least 10 cm$^{-3}$ must be present in the CGM of QSOs within a radial distance of 50 to 200 ckpc. To understand the origin of this discrepancy, we study the cold CGM density distribution finding in all cases simulated PDFs consistent with skewed lognormal ones with medians $\sim 10^{-2}$ cm$^{-3}$ and clumping factors between 16 and 102. We compare the broadness of these distributions to the one implied by the observed HeII/Ly$\alpha$ CGM ratio, which is independent on the assumption of maximum fluorescence. Assuming lognormal PDFs with median values similar to the simulations, the observed ratio requires a broader PDF with a clumping factor at least 8 times larger than predicted by simulations. These results suggest that higher resolution and/or additional physical mechanisms are needed to reproduce the observed CGM of QSOs at these redshifts.
}

\keywords{galaxies: halos -- galaxies: high-redshift -- galaxies: evolution -- quasars: general}
\maketitle


\section{Introduction}
Diffuse gas accreting onto galaxies constitute the reservoir for star formation and thus plays a central role in galaxy evolution. The way in which this accretion takes place is however still not fully understood. At $z \gtrsim 1.5$, the gas accreting along filaments is predicted to remain dense and cold ($<10^5$ K) until it reaches the central galaxy \citep{Fardal2001,Keres2005, Dekel2006,Agertz2009,Waterval:2025}. 
Insight on accretion mechanisms can thus be gained by studying the small-scale density distribution of the cold circumgalactic medium (CGM) at similar redshifts. 

The majority of observations of the CGM at high redshift are obtained through absorption features in the spectra of background quasars and galaxies. These observations have already placed numerous constrains on the state of the CGM but they only provide one dimensional information making it almost impossible to obtain complete information about the CGM density distribution and its morphology, necessitating its study in emission. Its low surface brightness (SB) at $z>2$ however makes it extremely challenging. Initial detections of the faint, extended Ly$\alpha$ emission surrounding galaxies where achieved by stacking large samples of narrow-band observations \citep{Steidel2011}, while Ly$\alpha$ nebulae are now ubiquitously detected around both star-forming galaxies and AGN at high redshift \citep{Cantalupo2014, Borisova2016, Wisotzki2016, Cai2017, Vanzella2017, Leclercq2017}. 
This development is largely due to new, more sensitive instrumentation, such as MUSE VLT \citep{Bacon2010}, which has the sensitivity necessary to detect emission as faint as $10^{-20}$ erg s$^{-1}$ cm$^{-2}$ arcsec$^{-2}$ in deep exposure times. 

There are three possible mechanisms responsible for powering the emission from Ly$\alpha$ nebulae: recombination radiation, collisional excitation, and ``continuum-pumping'' \citep{Cantalupo2017}. This makes deriving the physical properties of the CGM from the emission nontrivial as all three emission mechanisms depend on the presence of both neutral and ionized hydrogen, and all three mechanisms are likely to occur concurrently. Further complicating matters, due to the resonant nature of the Ly$\alpha$ line, the emission is subject to radiative transfer effects. To facilitate the physical interpretation of the Ly$\alpha$ emission, one can look at the Ly$\alpha$ nebulae associated with UV-bright quasars (QSO). The ionizing radiation from these bright QSOs is able to completely ionize the gas in the CGM on scales of hundreds of kpc \citep{Cantalupo2014, Umehata2019, Bacon2021}. The cold gas then emits photons through recombination radiation \citep{Haiman2001, Bunker2003, Cantalupo2005}. This quasar fluorescence boosts the emission from the CGM to SB values as high as $10^{-17}$ erg s$^{-1}$ cm$^{-2}$ arcsec$^{-2}$ at $z\geq2$, making the detection of extended Ly$\alpha$ emission easier \citep{Cantalupo2012, Cantalupo2014, Borisova2016, ArrigoniB2019, Cantalupo2019, Umehata2019, Fossati2021, Tornotti:2024}. 
The comparison of this resonant emission with non-resonant emission lines such as H-H$\alpha$ \citep{Leibler2018, Langen2023} and HeII-H$\alpha$ \citep{Cantalupo2019,Travascio:inprep} hints that recombination radiation is indeed the dominant emission mechanism for Ly$\alpha$ nebulae around bright QSOs. In the scenario of \textit{maximal fluorescence} (i.e. assuming that all the hydrogen is fully ionized by the QSO with a full opening angle) the emissivity of the gas is determined by its density squared and the temperature-dependent recombination coefficient. As the recombination coefficient is well constrained and varies weakly over the temperature range of the cold CGM ($10^4 - 10^5$ K), it is possible to constrain the density of the emitting gas from the observed SB maps. However, as observations currently still lack the resolution to resolve the structure of the cold CGM \citep{Crighton2015, ArrigoniBattaia2015, deBeer2023}, the SB also depends on the broadness of the cold gas density distribution (i.e., its clumpiness) within an individual resolution element, introducing an additional unconstrained variable. 

On the theoretical side, state-of-art simulations fail to numerically converge for the densities and scale of cold CGM sub-structures (e.g. \citealt{Hummels2019,Nelson2020,Benett:2020,RameshNeslon2023}). As with observations, it is thus not yet possible to obtain the precise physical properties of the cold CGM solely using simulations. This limitation is not solely due to resolution, but also to the difficulty of capturing the relevant physical dissipation scales. Complementary high-resolution idealized simulations can better resolve some of these processes, but typically rely on simplified geometries and remain sensitive to initial conditions  (e.g. \citealt{Mandelker2019,Ledos:2023,aung:2024,Yao:2025}). Despite their limitations, both recent zoom-in and high-resolution idealized simulations tend to indicate enhanced cold gas survival, driven by efficient cooling in the mixing layer between hot and cold gas (e.g. \citealt{Fielding2020}).  

The aim of the RePhyNe project (``Resolving the physics of Quasar Ly$\alpha$ Nebulae'') is to constrain the physical properties of the cold CGM around quasars by comparing observed Ly$\alpha$ nebulae associated with QSOs with simulated nebulae. However, due to the degeneracy between host halo mass, cold gas fraction, and density distribution of the emitting gas (clumping factor and mean density) on the SB values, performing this comparison is nontrivial. While it is possible to put constraints on the clumping factor with the non-resonant HeII-H$\alpha$ emission in the recombination dominated scenario \citep{Cantalupo2019,Travascio:inprep}, it is still necessary to first constrain the mass of QSO host halos. In the first paper of this series, \citet{deBeer2023} which we refer to as \citetalias{deBeer2023} in the remainder of this work, a new  mass estimation method was introduced, showing that QSO Ly$\alpha$ nebulae are typically hosted by halos within a mass range of $10^{12.03} - 10^{12.3} \mathrm{M}_{\odot}$ and no evolution with redshift. We consider here the slightly broader range $10^{12} - 10^{12.5} \mathrm{M}_{\odot}$ which is also consistent with other estimates in the relevant redshift range \citep{Shen2007, Trainor2012, FontRibera2013, Eftekharzadeh2015, garcia2017, He2017}. Armed with this estimate, in this second paper, we study the density distribution of the cold CGM. 

This paper is organized as follows: in \refsec{CosmSims}, we introduce our new simulations suite called {\sc DaLya}. In \refsec{methods} we present how we generate mock SB observations, and we recap some theoretical aspects of recombination SB. We present our results and compare them to observations in \refsec{results}. We start with the $\lya$ SB in \refsec{lya}, followed by the density distribution of the cold CGM in \refsec{denDistCGM}. We compare our results with those obtained from the state-of-the-art simulations HELLO \citep{Waterval:2024} and COLIBRE \citep{Schaye:2025,Chaikin:2025} in \refsec{simcomp}. Finally, we discuss our results in \refsec{discussion} and conclude in \refsec{conclusion}.

\section{The DaLya simulations} 
\label{sec:CosmSims}
\begin{figure*}
	\hspace{0.8cm}\includegraphics[width=1.9\columnwidth]{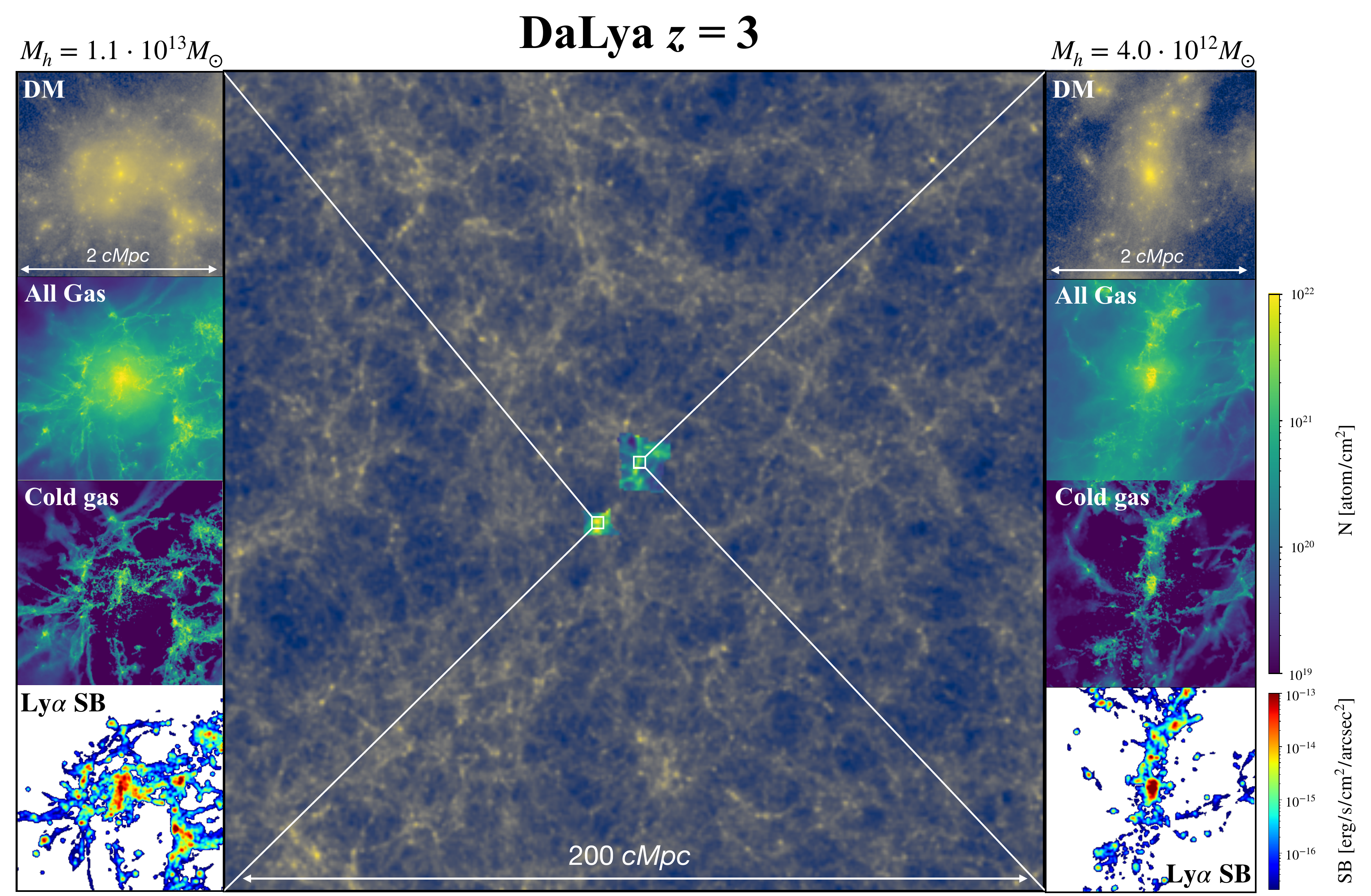}
	\caption{A visual impression of the {\sc DaLya} simulation (fiducial). The center panel shows the full base simulation with the two re-simulated regions highlighted. The inserts on the left and right present different quantities for two halos, one in each region, in a 2x2x2 cMpc volume. From top to bottom: projected dark matter density, gas column density, cold gas column density, and the $\lya$ nebula as extracted by \texttt{CubEx} \citep{Cantalupo2019} from the mock cube generated using the P2C output (see \refsec{MockOb} for details).}
	 \label{fig:zoomsim}	
\end{figure*}
The {\sc DaLya} (DArk matter and LYman-Alpha) simulations is a new suite of zoom-in hydrodynamic simulations ran with the public code {\sc Gizmo} \citep{Hopkins:2014qka}. The linear power spectrum was obtained from {\sc Camb} \citep{Lewis:1999bs}, the initial conditions (ICs) were generated at $z=99$ with {\sc MUSIC} \citep{Hahn:2011}, and we adopt the same cosmology as the Eagle simulations \citep{Schaye:2015}: $\Om=0.307$, $\OL=0.693$, $\Omega_b=0.0483$, $\OK=0$, $n_s=0.96$, $h=0.6777$. Halos were identified with a spherical overdensity algorithm using the public code AHF \citep{Gill:2004, Knollmann:2009}. The base simulation is a gravity-only simulation with a comoving box size $L=200$ cMpc and $N_p=128^3$ dark matter particles\footnote{Throughout the paper, we refer to particles in simulations even though most modern simulations do not have particles anymore to avoid any confusion with the pixels of the regular grids we use to build the mock observations.}, corresponding to a mass resolution of $m_p=1.5 \cdot 10^{11} M_\odot$. This large box size ensures the presence of rare massive halos with masses $\sim 10^{13}$ which are the focus of {\sc DaLya}. 

We identify two massive nodes in the base simulation corresponding to two protoclusters at redshift 3 with the most massive structure in each region having a mass of $7.2 \cdot 10^{12} M_\odot$ and $1.1 \cdot 10^{13} M_\odot$. These regions were selected as they are similar to the massive protocluster associated with MQN01 ($z=3.2$) observations \citep{Pensabene:2024, Galbiati:2024bsj, Travascio:2024,Pensabene:2025,Cantalupo:inprep}. We resimulate these massive nodes and their environment at two different resolutions adding baryons: the fiducial resolution (\dalya fiducial) corresponds to a dark matter mass of $m_{DM} = 3.8 \cdot 10^6 M_\odot$ and a mean baryonic mass of $m_{b} = 7.2 \cdot 10^5 M_\odot$. \dalya fiducial contains 12 halos in the mass range $10^{12}$ -- $10^{12.5}M_\odot$ at $z=3.5$ which is the focus of this work. We complete these with runs covering a smaller volume corresponding to $m_{DM} = 4.8 \cdot 10^5 M_\odot$ and $m_{b} = 9.0 \cdot 10^4 M_\odot$. We refer to these as \dalya HR. \dalya HR contains only two halos at $z=3.5$ but it allows us to study the impact of resolution at fixed subgrid physics. As a first visual impression, the projected density field of the base simulation with insets corresponding to part of the two zoom regions at the fiducial resolution are shown in \reffig{zoomsim}. We show the projected DM density (top), total and cold gas only column density (center), and the associated $\lya$ emission (bottom, see \refsec{MockOb}). 

We perform adaptive gravitational softening with a standard value of 32 neighbors for both dark matter and gas particles. For dark matter, we impose a minimum smoothing length $\epsilon_{DM} = 1.2$ ckpc roughly corresponding to 1/40$^{\rm th}$ of the mean interparticle distance. For gas particles, we must ensure that we resolve high enough densities corresponding to the star formation criterion we adopt (see \refsec{hydro}). We compute it as $\epsilon=H/1.4$ with $H$ the mean interparticle distance of a sphere of density corresponding to $n_{\rm thresh}$ (the threshold for star formation) containing 32 neighbors. For our choice of $n_{\rm thresh}$ this gives a minimum smoothing length $\epsilon_{gas} = 0.1$ ckpc in \dalya fiducial, and $\epsilon_{gas} = 0.05$ ckpc in \dalya HR.

\subsection{Hydrodynamical model}
\label{sec:hydro}

We now turn to a detailed description of the hydrodynamic model. We calibrate our model to reproduce the stellar-to-halo mass function at $z=3$ as well as the neutral hydrogen column density PDF at $z \gtrsim 3$. The quantitative validation of the model is presented in \refapp{dalya}.

We use the Meshless Finite Mass (MFM) method implemented in {\sc Gizmo} to solve the hydrodynamic equations. We adopt a standard polytropic equation of state index $\gamma_{\rm poly}=5/3$. Galactic star formation is triggered above a physical density threshold of $n_{\rm thresh} = 10$ cm$^{-3}$. Above this threshold, gas particles are turned into stellar ones on a probabilistic basis following a smooth rate proportional to $m_g/t_{dyn}$, where $t_{dyn}$ is the particle dynamical time \citep{Springel:2002uv}. Stellar particles reflect entire initial mass function (IMF)-averaged populations for which we use the IMF from \cite{Chabrier:2003}. 

A fraction of stars result in supernovae (SN). This is the only source of feedback since we do not include black hole seeding or feedback from massive (AGB) stars. The SN feedback model is based on the FIRE and FIRE-2 simulations \citep{Hopkins:2014fire, Hopkins:2017ycn, Hopkins:2018sn}: we assume an age-dependent SN rate $R$ of 
\begin{equation}
R = \Bigg\{ \begin{array}{ll}
5.408 \cdot 10^{-4} \;\; {\rm if} \;\; 3.4 < {\rm age} < 10.37,\\
2.516 \cdot 10^{-4} \;\; {\rm if} \;\; 10.37 < {\rm age} < 37.53,\\
0 \;\; {\rm if} \;\; {\rm age} > 37.53,
\end{array}
\end{equation}
where rates are in SNe/Myr/$M_\odot$ and the age is in Myr. At each time step, for each stellar particle, the probability $p$ of an event (Type-Ia and/or Type-II) is determined based on the mass, age, and metallicity. The feedback itself is both thermal (energy is injected in nearby particles) and mechanical (mass transport through winds). Our model assumes that mass ejecta have a velocity of 4377 km/s, corresponding to $2\cdot 10^{51}$ erg for a mass of $10.5 M_\odot$. Note that the FIRE-2 values are set to $10^{51}$ erg per SN and an ejecta speed of 3095 km/s. We adopt slightly higher values to compensate the fact that we do not model AGN feedback and avoid overcooling, thus matching the stellar mass to halo mass relation (\reffig{calibration}). We refer the reader to \cite{Hopkins:2014fire, Hopkins:2017ycn,Hopkins:2018sn, Lupi:2018} for a  detailed discussion of SN feedback in the FIRE and FIRE-2 simulations.

In addition to feedback, supernovae are also responsible for metal enrichment  (e.g. \citealt{Oppenheimer:2008,Shen:2012}). As in FIRE-2, we follow 11 atomic species, from hydrogen (H) and helium (He), to iron (Fe) \citep{Hopkins:2017ycn}. Primordial abundances are all set to zero in the ICs except for hydrogen and helium. The cooling routine computes the total cooling rate from the abundance of each element for temperature down to $T= 10^4$ K following tabulated values computed with CLOUDY in \cite{Wiersma:2009}. Fine-structure and molecular cooling are also added to extend the cooling curve down to $10K$ following \cite{Hopkins:2017ycn, Hopkins:2022qzm}. These cooling terms are only relevant for high densities (above 1 atom/$cm^3$) as can be seen in \reffig{Trhofull}, and are also counter-balanced by heating from cosmic rays. The complete list of all cooling physics included is given in Appendix B of \cite{Hopkins:2017ycn} (FIRE-2), while more recent updates are given in section 3 of \cite{Hopkins:2022qzm} (FIRE-3). 

Photo-heating and photoionization (reionization) are implemented through a spatially uniform UV background. We use tabulated values from \cite{Faucher-Giguere:2019kbp} appropriately corrected following \cite{Onorbe:2016hjn} to reionize at the correct target redshift, and taking into account the updated AGN contribution following \cite{Shen:2020obl}. We include hydrogen self-shielding from the UVB following \cite{Rahmati:2013tr, Hopkins:2022qzm}.  

\section{The CGM in emission} 
\label{sec:methods}

We now turn to how we produce mock $\lya$ observations from simulations outputs, as well as theoretical expectations for the dependence of SB values on the cold gas properties.

\subsection{Mock integral-field-spectroscopy observations} 
\label{sec:MockOb}
We generate $\lya$ mock observations under the assumption of maximal fluorescence, i.e. assuming that all the hydrogen in the simulation is fully ionized by a sufficiently bright quasar with a full opening angle. In this assumption, Ly$\alpha$ emission is generated by recombination only. This is a reasonable assumption for the CGM of the bright QSOs which are the subject of the present study \citep{Leibler2018, Cantalupo2019, Langen2023}. For simplicity, we also neglect the effects of local Ly$\alpha$ radiation scattering due to residual neutral hydrogen in dense regions. A proper modeling of this effect would require radiative transfer models performed at a resolution which is likely currently not achieved by cosmological simulations. We therefore calculate the gas Ly$\alpha$ emissivity $\epsilon_{\mathrm{Ly}\alpha}$ using the relation
\begin{equation}
    \epsilon_{\mathrm{Ly}\alpha} = \frac{1 - Y/2}{1-Y}\frac{h\nu_{\mathrm{Ly}\alpha}}{4\pi} n^2\alpha_{\mathrm{eff}}(T) \, ,
    \label{eq:LyaEm}
\end{equation}
where $n$ is the hydrogen number density, $\nu_{\mathrm{Ly}\alpha}=1215.67\mathring{A}$ is the Ly$\alpha$ rest frame frequency, $h$ is Planck's constant, and $(1-Y/2)/(1-Y)$ is a correction term due to primordial helium, where $Y$ is the primordial helium's number fraction. The case A effective recombination coefficient for Ly$\alpha$, $\alpha_{\mathrm{eff}}(T)$, is based on the fraction of recombination events including the ground state that result in Ly$\alpha$ emission as calculated by \citet{Cantalupo2005,HuiGnedin1997}.

In addition to ionizing the CGM, the central quasar also heats the surrounding medium through photo-electric heating. Due to the gas cooling when it recombines and emits radiation, the temperature of the gas will eventually reach an equilibrium between photo-heating and cooling. We refer to this equilibrium temperature as the gas photo-heating floor. To account for this effect, we impose a photo-heating floor of $5 \times 10^4$ K, as calculated in \citet{PezzulliCantalupo2019} for an ionized medium with a number density of 1 cm$^{-3}$ and a metallicity of 0.1 $Z_{\odot}$ at a distance of 50 kpc from the central, ionizing quasar, assuming a standard quasar spectrum \citep{Lusso2015}. We have explicitly checked that our results and conclusions are unaffected by the exact choice of photo-heating floor. We emphasize that we do not model the hydrodynamic response of the gas to such heating, i.e we do not modify the gas density. Given this lack of hydrodynamical response and the assumption of maximal fluorescence, the modeled SB values in our mock observations should be considered as strict upper limits, in the absence of other Ly$\alpha$ emission mechanisms than recombination radiation.  

\begin{figure}
	\centering
    \includegraphics[scale=0.35]{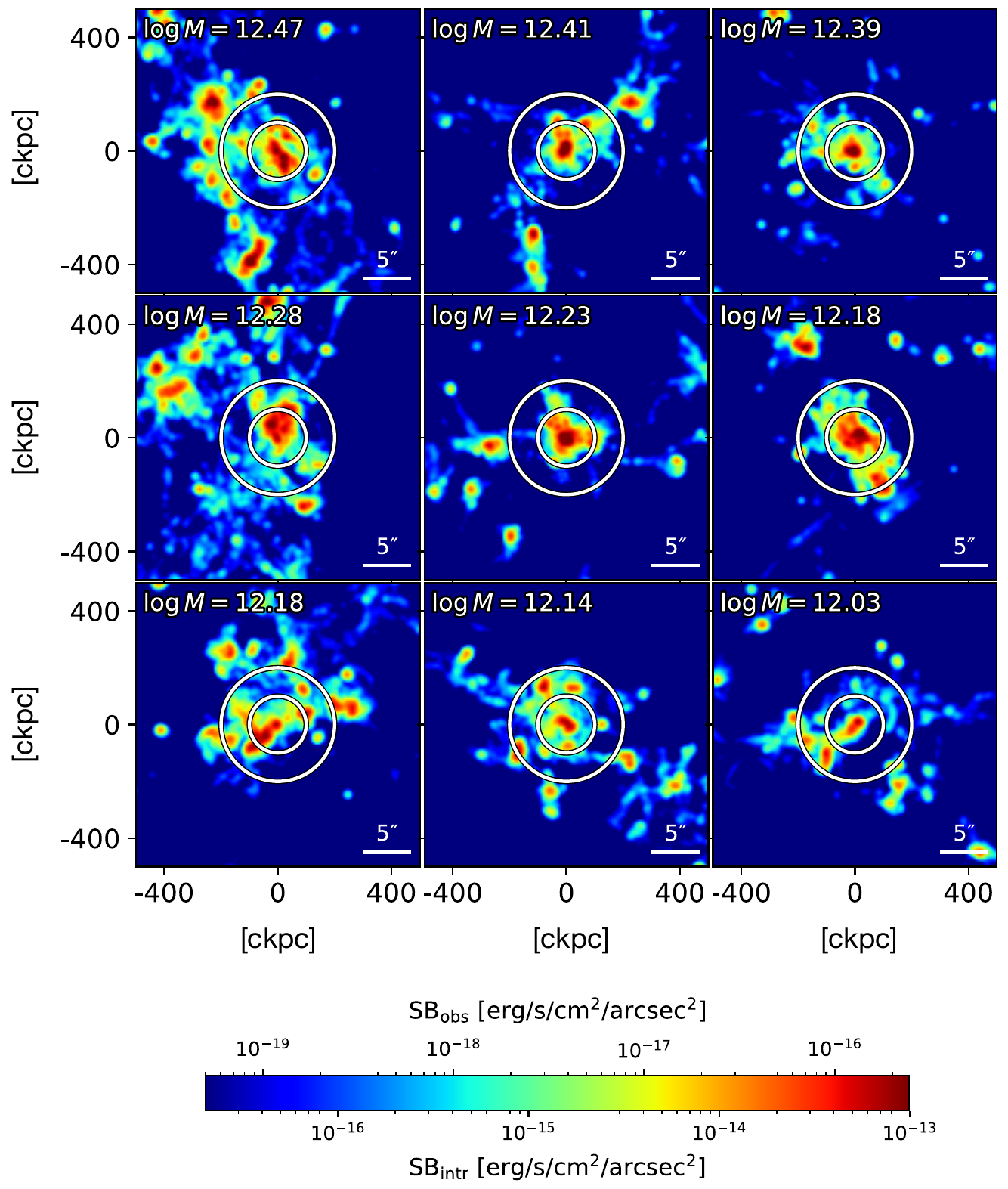}
    \caption{Mosaic of selected mock pseudo-narrow band $\lya$ SB maps (without noise for clarity) at $z=3.5$ in a 1$\times$1 cMpc field of view for \dalya fiducial. Each panel shows a randomly selected LOS for halos spanning the mass range $12.0 \leq \log M \leq 12.5$. The white circles indicate the radial annulus 100 -- 200 ckpc.}
    \label{fig:mosaicSB}
\end{figure}

We generate MUSE compatible mock integral-field-spectroscopy observations using the procedure described in detail in \citetalias{deBeer2023}. For each halo, we populate a grid of voxels with the emissivity of all gas particles within a $2\times2\times2$ cMpc$^3$ box centered on the halo. Notice that doing so we assume a QSO opening angle of 100\%. When populating the grid, we also remove the interstellar medium (ISM, i.e. star forming particles) associated to satellite galaxies, which should remain self-shielded even in the presence of a bright quasar and for which Ly$\alpha$ dust attenuation would be significant. This is done based on the same criterion as in the simulations, i.e. by ignoring the contribution from particles denser than 10 atoms per cm$^{3}$. 

We then convert the grid to mock cubes based on the position and mean velocity of the voxels perpendicular to the line-of-sight (LOS). The spatial and spectral resolution of the mock cubes are chosen to be as close as possible to that of the MUSE integral field spectrograph, which has a spatial sampling of $0.2 \times 0.2$ arcsec$^2$ and a spectral resolution of 1.25$\mathring{A}$. To mimic more closely the typical observing conditions, we perform an extra smoothing on a scale of 1.5 pixel (corresponding to a FWHM of 0.8 arcsec or 27.5 ckpc), and we add artificial noise spectral layer by spectral layer. The noise has a Gaussian distribution with a standard deviation of $5\times10^{-20}\,\, \mathrm{erg}/(\mathrm{s} \,\, \mathrm{cm}^2 \, \mathrm{arcsec}^2 \, \si{\angstrom})$ and corresponds to the noise level achieved with integration times of 10 hours or longer with MUSE.  We then extract the Ly$\alpha$ nebulae from the mock cubes using \texttt{CubEx} from the \texttt{CubeExtractor} package \citep{Cantalupo2019} applying parameters that are consistent with those used on actual observations. We stress that the exact value of the applied artificial noise does not affect our results and its only purpose is to extract mock nebular emission in a similar fashion to observations. Since the extraction depends on the SNR ratio and our nebulae are typically fainter than observations (as discussed below), we use a conservative low value of artificial noise to extract nebulae which are large enough to be compared to observations. In particular, for a nebula to be extracted by \texttt{CubEx}, each of its voxels must have a signal-to-noise ratio higher than two, and it must consist of more than 1000 voxels. We then create pseudo-narrow-band (NB) images by considering the most extended object extracted by \texttt{CubEx} after verifying that it coincides with the center of the halo hosting our assumed ionizing source, and summing voxels along a line-of-sight from the smallest to the highest spectral layer in which the nebula is detected (i.e. all pixels in the 2D image result from the sum along the same spectral range). Finally, we produce three mock cubes along three perpendicular LOS for each halo. 

By populating the grid with particles and then integrating over the LOS we introduce a clumping factor along the LOS to the mock observations. The clumping factor of a pixel quantifies the variation of the density along the LOS, as well as the unresolved broadness of the gas density distribution within that pixel (subresolution clumping factor), and is defined as
\begin{equation}
    C \equiv \, \left\langle n^2\right\rangle_l / \left\langle n\right\rangle_l ^2.
\label{eq:clumping}
\end{equation}
The spatial scale or volume over which the integration is performed is $l$ and $n$ refers to the gas density. If the density of the gas is constant on $l$, the clumping factor is equal to one by definition and greater than one otherwise. Although the clumping factor of the real, observed structures is also defined in this way, it is important to note the actual values of the mock observation's clumping factor and that of the observations are different: as the simulations do not completely resolve the CGM, they also do not model the gas's entire density distribution, leading to a different value with regard to observations for the clumping factor.   

The bottom insets in \reffig{zoomsim} give a visual impression of the mock $\lya$ nebulae corresponding to the selected systems with noise subtracted, showing that it primarily follows the cold dense gas in filaments. \refFig{mosaicSB} further shows a more extended subsamble of noise-subtracted pseudo-NB images over $1\times 1$ cMpc$^2$ field of view. Each time, the halo center (thus the respective QSO) is at (0,0). We show 9 LOS from 8 different halos for \dalya fiducial. The two circles in each panel show the 100 -- 200 ckpc annulus which we use later to constrain CGM properties (\refsec{results}). We see that the extension of the emission diminishes with decreasing mass, and that it is far from spherically symmetric in most cases. \refApp{addmosaic} presents additional SB maps for \dalya HR, as well as the COLIBRE (m6 and m5) and HELLO simulations with which we compare our results in \refsec{simcomp}.
 
\subsection{Fundamentals of recombination surface brightness}
\label{sec:theorExp}

As a reference for interpretation later on, we recall here some basic properties of the recombination radiation SB. For a fully ionized gas, a given line of sight and neglecting radiative transfer effects, the Ly$\alpha$ SB due to recombination is given by
\begin{equation}\label{eq:basic}
{\rm SB} = \int B(T(l))\, (n(l))^2\, dl \; ,
\end{equation}
where $l$ is a coordinate along the line of sight, $n$ is the hydrogen number density, while the coefficient $B(T)$, defined in equation (2) of \cite{PezzulliCantalupo2019}, is directly proportional to the effective Ly$\alpha$ recombination coefficient and thus a declining function of $T$.

If the temperature of the gas along the line of sight is approximately constant, then \refeq{basic} can also be written
\begin{equation}
{\rm SB} = B(T)\,  C \left\langle n \right\rangle^2 \; ,
\end{equation}
where the (global) clumping factor $C$ is defined in \refeq{clumping} and the averages $\left\langle \cdot \right\rangle$ are computed along the whole line of sight (see e.g. \citealt{Cantalupo2014}).

In many applications, however, and especially for the CGM of massive
halos, $T$ is very far from constant, with most of the volume occupied
by gas so hot ($T> 10^6$ K) that $B(T)$ is negligibly small, while
virtually all the recombination radiation is emitted by a small fraction
of the volume (defined as the volume filling factor $f_V \ll 1$)
occupied by cold and dense gas with instead approximately constant (and
lower) temperature $T$ (typically, $10^4 < T/\textrm{K} < 10^5$). Under
this refined approximation, \refeq{basic} then becomes
\begin{equation}
{\rm SB} = B(T) f_V \, C_\textrm{cold} \left\langle n \right\rangle_\textrm{cold}^2 \; ,
\label{eq:SBCfv}
\end{equation}
where, this time, the averages $\left\langle \cdot \right\rangle_\textrm{cold}$ are computed only within the minority of the volume occupied by the cold gas.

We recall (see also Section 2.1.2 in \citealt{PezzulliCantalupo2019}) that, in the same approximation, the global clumping factor (\refeq{clumping}) can
also be written as
\begin{equation}
C = C_\textrm{cold}/f_V \; .
\end{equation}
For example, a filling factor $f_V \sim 10^{-3}$ could already explain a
global clumping factor $C \sim 1000$ even if the cold gas had a single density. At the same time, for a fixed value of $f_V$, density variations within the cold gas itself (parametrized by $C_\textrm{cold}$) can further increase the value of $C$, therefore affecting the final value of the surface brightness (see also \citealt{PezzulliCantalupo2019};
\citealt{Cantalupo2019}).

\section{Results}
\label{sec:results}

\subsection{Ly$\alpha$ \hspace{1pt} surface brightness} 
\label{sec:lya}

\begin{figure*}
	\centering
	\includegraphics[scale=0.39]{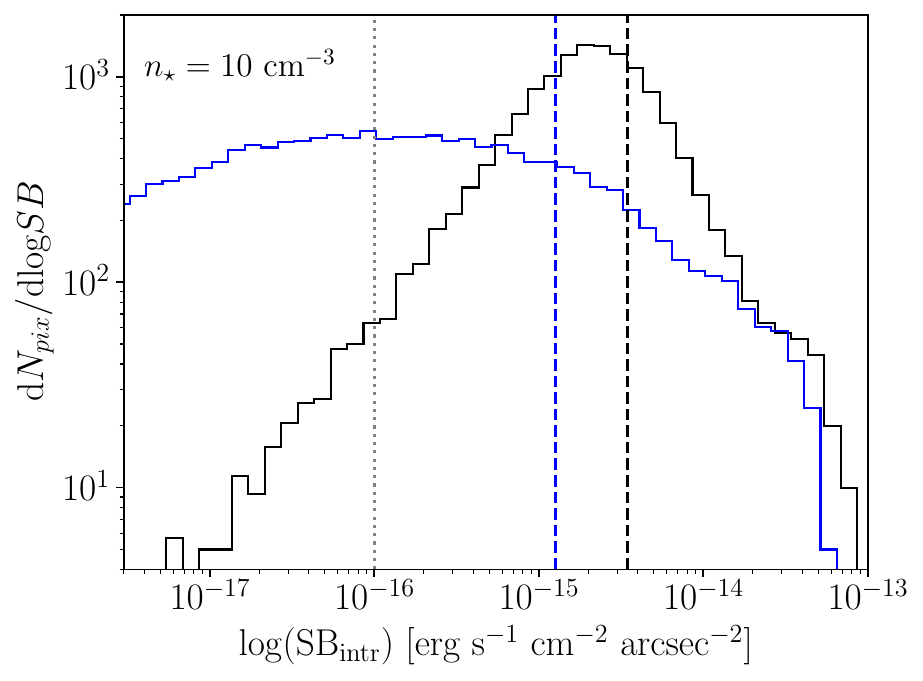}
    \hspace{-0.3cm}\raisebox{0.0cm}{
    \includegraphics[scale=0.39]{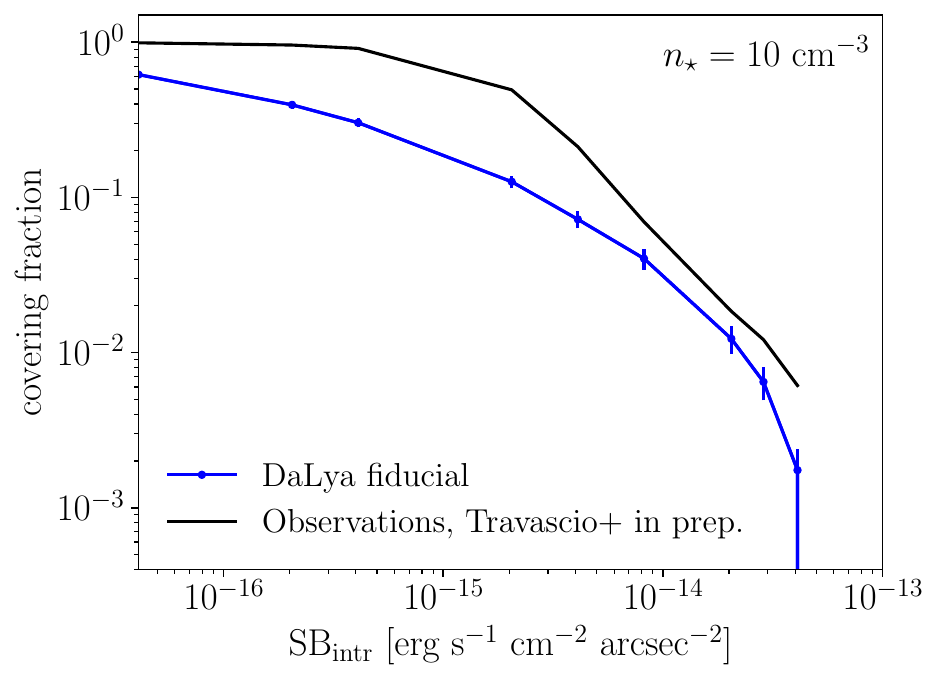}}
    \hspace{-0.2cm}\includegraphics[scale=0.39]{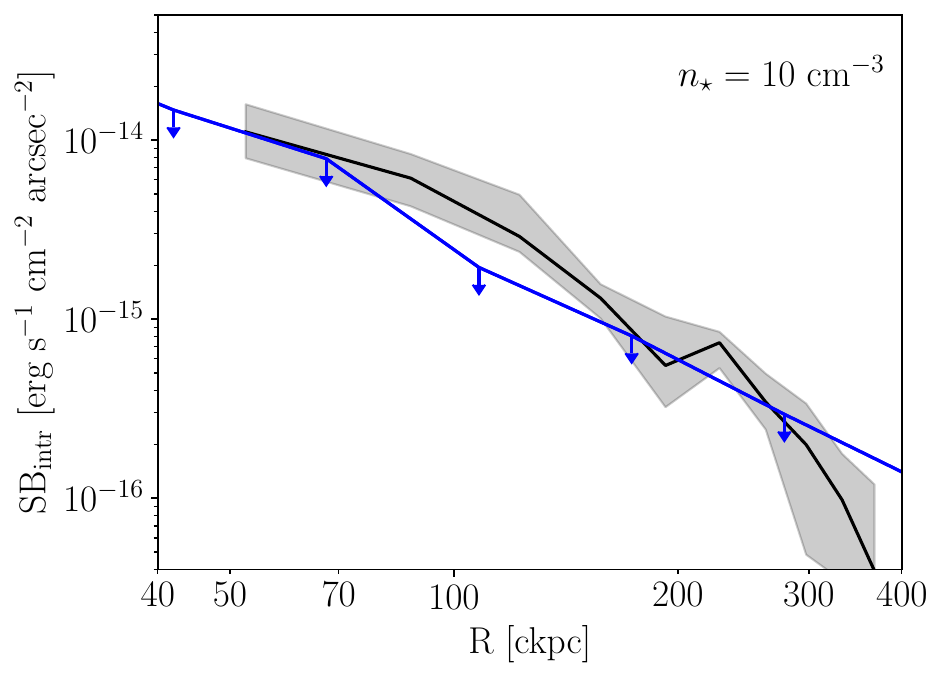}
    \caption{\textbf{Left panel:} The $\lya$ SB function (i.e. the count of intrinsic SB values per logarithmic SB bins) at $z=3.5$ in the 100-200 ckpc annulus and comparison with observations from \citetalias{Travascio:inprep}. Results for \dalya fiducial are shown in blue while results from observations are in black. Each time we show the mean number of pixels per LOS. The dashed vertical lines indicate the mean of each histogram, and the dotted vertical line indicates the average detection limit in observations. \textbf{Central panel:} The covering fraction of $\lya$ pixels, corresponding to the cumulative version of the histograms on the left panel. This representation highlights the very different morphology of the emission between \dalya and observations as simulated nebulae never reach a covering fraction of 1. \textbf{Right panel:} Circularly-average SB profiles, corresponding to the median of the circularly averaged profiles of each halo (and 3 different LOS). We omit the error on the simulated profiles for clarity and use arrows to emphasis that these profiles correspond to upper limits within our assumptions. As discussed in \refsec{results}, since these profiles are based on averages, they cannot capture the discrepancies in the detailed SB distribution properties, and thus on the emitting gas density distribution.}
    \label{fig:SB100200z3p5}
\end{figure*}

We now present our results for the $\lya$ SB in \dalya fiducial. For the remaining of the paper, we fix the redshift of simulations to $z=3.5$, and the halo mass range to $12.0 \leq \log_{10} M \leq 12.5$. All results are obtained from mock pseudo NB observations assuming maximum fluorescence, and excluding ISM particles (i.e. particles with density above 10 cm$^{-3}$) following the procedure described in \refsec{MockOb}. Finally, we show intrinsic $\lya$ SB values to remove redshift-dimming effects and make the comparison between simulations and observations more fair, as intrinsic SB values do not show strong variations with redshift around $z \sim 3.5$ \citep{Kusabe:2022,Niemeyer:2022}.

We compare our results with observations of 14 quasar Ly$\alpha$ nebulae obtained with MUSE on fields centered on bright radio-quiet QSOs at median redshift $z=3.76$ selected from the MAGG and MUSEQuBES surveys \citep{Lofthouse2020, Fossati2021, Muzahid:2020, Muzahid:2021}. We choose this sample as, in addition to $\lya$ emission, these nebulae also present individual detection of extended HeII-H$\alpha$ emission which we will use to place constraints on the density distribution in the cold CGM, and compare with what is obtained in simulations (\refsec{denDistCGM} and \ref{sec:simcomp}). These observations are presented in detail in a companion paper \citetalias{Travascio:inprep}, and are simply denoted as ``observations'' in the following. We have explicitly verified that the $\lya$ SB values of this particular subsample are consistent with their parent samples and the largest sample of quasar Ly$\alpha$ nebulae available from the the literature \citep{Borisova2016, ArrigoniB2019, Fossati2021}. As such, we can consider these observations as representing the typical Ly$\alpha$ SB of bright QSOs at this redshift.

In the left panel of \reffig{SB100200z3p5}, we first investigate the average number of pixels per logarithmic SB bin, ${\rm d} N_{\rm pix} /{\rm d} \log$SB, in the 100 -- 200 ckpc radial annulus. In analogy with the halo mass function, we refer to this quantity as the \textit{$\lya$ SB function}. Results for {\sc DaLya} are presented in blue while results from observations are shown in black.

In both cases, the $\lya$ SB function displays a quasi lognormal behavior with a tail towards low SB values. However, the simulated curve is broader and peaks at a lower value than the observed one. We note that the peak of the observed $\lya$ SB function is more than one order of magnitude above the detection limit situated on average at 10$^{-16}$ erg/s/cm$^2$/arcsec$^2$ (dotted grey line). As such, the different shapes of the Ly$\alpha$ SB functions of observations and \dalya cannot be ascribed to observational limitations and, within our assumptions, must thus reflect differences in the underlying gas distribution. This is also confirmed by the $\lya$ covering fraction in the same radial annulus, presented in the middle panel of \reffig{SB100200z3p5}, equivalent to the cumulative $\lya$ SB functions. There are few bright pixels (with SB $> 10^{-14}$) in observations, but the curve then rises quickly and reaches a covering fraction of 100\% at $5\times 10^{-16}$. This shows that, on average, within the 100-200 ckpc annulus, all pixels are detected as $\lya$ nebulae in observations. This is in contrast with \dalya where the fraction never reaches 1. This is also visible in \reffig{mosaicSB} where numerous non-detected pixels (dark blue) are present within the annuli delimited by the white circles. 

To our knowledge, this is the first time that such comparison has been made. Previous observational and theoretical studies characterized the Ly$\alpha$ SB within a given annulus with a single value, e.g. the mean, which cannot fully capture the discrepancies shown in \reffig{SB100200z3p5} due to the fact that such quantity is dominated by the brightest pixels. Indeed, as shown by the dashed vertical lines in the left panel of \reffig{SB100200z3p5}, the observed mean Ly$\alpha$ SB is within a factor of 3 with of the simulated one. This is further  illustrated in the right panel of \reffig{SB100200z3p5} which presents the SB profile for the radial range 40 -- 400 ckpc. As commonly done, we present the median of the circularly averaged profiles for each halo (and 3 different LOS). We omit errorbars on the simulated profiles for clarity and use down pointing arrows to emphasis that these profiles are upper limits within our maximal fluorescence assumptions, which also include the assumption of a 100\% opening angle for the quasar radiation and no hydrodynamical response of the gas. Despite the differences in the $\lya$ SB function, the simulated profile is within a factor of 3 from the observed one over the full radial range. Our work thus highlights that caution should be exercised when interpreting results based on averaged quantities. Since the simulated profile is an upper limit, the relatively good agreement with the observed one implies that densities of at least 10 cm$^{-3}$ must be present in the the cold CGM of bright QSOs at $z\sim3.5$.

\subsection{Density distribution of the cold CGM}
\label{sec:denDistCGM}

\begin{figure}
	\centering
	\includegraphics[scale=0.52]{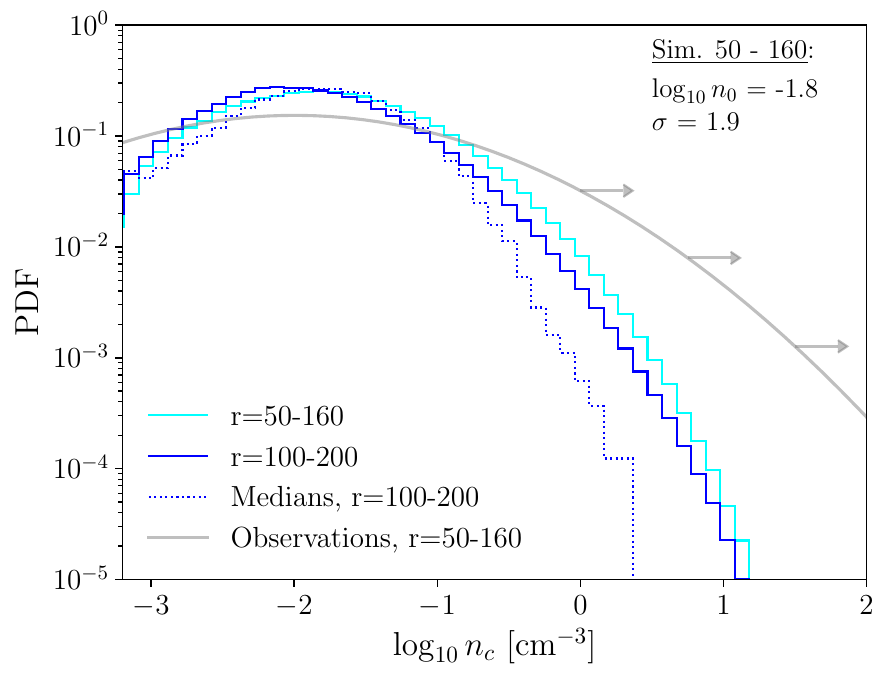}
    \caption{Average PDFs of the density of the cold ($10^3 \leq T \leq 10^5$K) CGM in \dalya fiducial at $z=3.5$ projected in two radial annuli. The solid histograms show the density distributions of simulations particles. For the 50-160ckpc annulus, we quote the median density $n_0$ and the 1$\sigma$ equivalent width of the distribution of ln$n_c$ (see \refapp{skewed}). The solid gray line shows the inferred lognormal PDF from observations of the HeII/$\lya$ ratio from \citetalias{Travascio:inprep} in the 50-160ckpc radial bin when fixing $\log_{10}n_0=-2$ and the corresponding minimum width $\sigma=2.6$ compatible with the observed ratio. The grey arrows illustrate that this distribution is a lower limit. The dotted blue histogram shows the distribution of medians of all individual PDFs along the LOS in each P2C pixels as discussed in \refsec{denDistCGM}.}
    \label{fig:DensityPDF}
\end{figure}

In the maximum fluorescence limit, the $\lya$ SB is directly proportional to the cold gas mean density square times the clumping factor times the volume filling factor (\refeq{SBCfv}). The discrepancy between the observed and simulated $\lya$ SB function may thus point towards a discrepancy in the underlying density distribution. We investigate the distribution of volume-weighted densities in \reffig{DensityPDF}. We select the gas in the temperature range $10^3$ -- $10^5$K (before QSO photo-heating) which is cold enough to be $\lya$ emitting and avoid the very cold molecular phase which would be affected by dust and for which our maximal fluorescent approximation does not hold. Importantly, we do not impose any density threshold. The volume of each particle is obtained by dividing the particle's mass by its density. We show results for particles whose projected positions fall within the 100 -- 200 ckpc annulus (blue) to allow direct comparison with results for the SB values, as well as within the 50 -- 160 ckpc one (cyan) for which we have constraints on the density distribution from the observed HeII/$\lya$ ratio from \citetalias{Travascio:inprep}, assuming that the emission from the two lines is co-spatial and that they probe the same gas, as discussed in \citetalias{Travascio:inprep}.  

The histograms display quasi-lognormal distributions skewed towards higher densities. While the density distribution of the ISM has been studied and is known to follow a skewed lognormal (e.g. \citealt{Basu2015}), this is the first time to our knowledge that it is systematically studied for the CGM. According to the central limit theorem, a lognormal density distribution is the natural outcome of multiplicative, random processes, making it a good starting point to describe the density distribution of a medium which is most likely subject to instabilities and turbulent processes \citep{Vossberg2019, Gronke2022}. While the central limit theorem is what governs the starting distribution of the masses, interactions with the surrounding medium over varying timescales skew the distribution and lead to a power law at higher densities \citep{Basu2004, Basu2015, Padnos2018, Vossberg2019, Mandelker2019, Mandelker2019ApJ, Gronke2022}. We discuss in more details the shape of the density PDFs in \refapp{skewed}.

We also quote the median density $n_0$ as well as the 1$\sigma$ equivalent width of the $\ln n_c$ distribution obtained from the 16$^{\rm th}$ and 84$^{\rm th}$ percentiles. We quote the width of the natural logarithm distribution to more directly relate it to the $\sigma$ parameter entering (skewed) lognormal distributions (see \refapp{skewed}). In the 50 -- 160 ckpc annulus, we find a median log$_{10} n_c$ of -1.8 and $\sigma=1.9$, corresponding to a clumping factor $C_{\rm cold} \approx25$. We emphasis that this is the clumping factor of the measured distribution and not the subresolution clumping factor which is not measurable. It is thus not fully equivalent to the definition in \refeq{clumping} for observations, but gives us an estimate of the amplitude of the effect. We also emphasis that the widths of the distributions shown in \reffig{DensityPDF} are also determined by the combination of additional factors, mainly the radial density gradient, the density gradient along the LOS, azimuthal asymmetries at fixed radius, and halo-to-halo variations. 

As shown in \cite{Cantalupo2019}, the HeII/Ly$\alpha$ line ratios of QSO Ly$\alpha$ nebulae can be used to constrain the underlying density distribution in the assumption of recombination radiation and for a given quasar spectrum. By using this methodology, and by assuming a lognormal distribution for the gas PDF, \citetalias{Travascio:inprep} have derived the inferred PDF broadness from QSO Ly$\alpha$ nebulae at $z\simeq3.8$ which can be directly compared to our results. In more details, they use CLOUDY \citep{Ferland2013} to model a cloud with size 200 pc at a distance 50--160 ckpc from a QSO with a soft UV spectrum slope $\alpha_{UV}=-1.7$ (compatible with results in \citeauthor{Lusso2015} \citeyear{Lusso2015}) and a minimum luminosity $\log(\nu L_{\nu} / \text{erg\,s}^{-1}) = 46.6$ at $1450~\AA$. Assuming that each cloud receives the full unattenuated HeII-ionizing photon flux, and fixing the median density to $\log_{10} n_0 = -2.0$, they find $2.6 \leq \sigma \leq 3.8$ to match the observed HeII/$\lya$ values, which is substantially higher than in {\sc DaLya}. The PDF with the minimum $\sigma$ compatible with observed line ratios is shown in grey in \reffig{DensityPDF}. As the clumping factor is roughly proportional to $e^{\sigma^2}$ (see \refapp{skewed}), even small differences in sigma represent orders of magnitude of difference between the simulated and observed clumping factors, as is shown in \reffig{clumping} for lognormal and skewed lognormal distributions. In our case, the difference in $\sigma$ translates into a clumping factor $\sim 35$ times larger in observations than in simulations. Notice that fixing the value of $\log_{10}n_0 = -2$ is arbitrary (as the mean density of the CGM is unknown), and that fixing it to a higher value would result in a lower $\sigma$. We note however that fixing $\sigma =1.9$ as suggested by our simulations leads to $\log_{10}n_0 > -1$ to match the observations, and a very high mean density $\left\langle n_c \right \rangle > 10$ cm$^{-3}$.  

Finally, we also show the histogram of the median values of the density distribution in each P2C pixel (corresponding to the same size of MUSE pixels) for the 100 -- 200 ckpc annulus: for each pixel, we consider all particles contributing to the emission of that pixel, compute the median density of those particles, and plot the distribution of these medians. The result is depicted by the dotted histogram in \reffig{DensityPDF}. As one can expect, the peak of the distribution slightly shifts towards higher values with respect to the histograms based on simulations particles, and the histograms become narrower. The histograms of median values illustrate an intermediate step between the density PDFs of all particles, and the $\lya$ SB histogram presented in \refsec{lya}: while the emissivity in each pixel is obtained from the individual particle densities, the computed SB values are then smoothed on the pixel scale (at minimum, if no other smoothing is applied). This smooths out high SB values in the same way as high densities are suppressed in the distribution of medians.

\subsection{Comparison with other simulations} 
\label{sec:simcomp}

We now compare the results presented so far to other simulations at the same resolution as \dalya fiducial, as well as at 8 times higher resolution, summarized in \reftab{sims}. We first compare with \dalya HR to study the impact of resolution at fixed subgrid physics. Second, we use the new publicly available COLIBRE simulations \citep{Schaye:2025, Chaikin:2025}, in particular the L050 m6 and m5 which have resolution matching \dalya fiducial and HR respectively. Finally, we use a subsample of the HELLO project introduced in \cite{Waterval:2024} with the same resolution as \dalya HR. In particular, we use halos in the g9.20e12, g2.71e12 and g2.32e12 grids. 

COLIBRE is a suite of cosmological simulations ran with the public code {\sc Swift} \citep{Schaller:2024} with additional subgrid modules. The hydrodynamics equations are solved with SPH. The number of dark matter particles is four times that of baryonic ones which suppresses spurious transfer of energy from dark matter to baryons and hence improves the effective resolution for baryons (e.g. \citealt{Ludlow:2023}). Radiative cooling is allowed down to $\approx 10$K \citep{Ploeckinger:2025} thus directly modeling the thermal balance in the ISM instead of imposing an equation of state. Cooling rates are computed from a CHIMES network following 11 elements, treating hydrogen and helium fully in non-equilibrium. The local radiation field is computed using a homogeneous UVB, an interstellar radiation field, and cosmic rays. Self-shielding by gas and dust is also implemented. The shielding column density is computed from the local Jeans length. The dust abundance is computed on the fly following \cite{Trayford:2026}, and is coupled to cooling. In contrast with \dalya and HELLO, star formation is not triggered above a given density threshold, but it is implemented using a Schmidt law combined with a gravitational instability criterion \citep{Nobels:2024}. The chemical enrichment due to AGB stars, massive stars, core-collapse SN, and type Ia SN is computed following \cite{Correa:2026}. Feedback processes include pre-SN feedback from massive stars \citep{Benitez-LLambay:2026}, core-collapse SN \citep{Chaikin:2023,Schaye:2025}, and SN Ia. We use the simulations implementing thermal AGN feedback \citep{Booth:2009,Schaye:2025} but we note that simulations implementing the hybrid model from \cite{Husko:2026} are also available. 

HELLO are zoom-in simulations of massive halos using the SPH code {\sc Gasoline2} \citep{Wadsely:2017}. Star formation is allowed above a density threshold of 10 cm$^{-3}$ and below a temperature of $1.5\cdot 10^4 K$. The feedback implementation includes stellar feedback from young bright stars, stellar winds, and SNe, as well as thermal AGN feedback. The chemical abundance of the 10 most abundant elements are computed. Finally, in addition to a homogeneous UVB, photoionization and photo-heating from three types of local radiation sources are implemented. Crucially, no self-shielding from the homogeneous UVB is performed. For further details of COLIBRE and HELLO, we refer the reader to the original articles. 
\begin{table}
\centering
  \begin{tabular}{ c | c | c | c }
    \hline
    Name & $m_{DM}$ & $\bar{m}_b$ & $\epsilon_b$ \\ \hline
    [-] & [$10^5 M_\odot$] &  \rule{0pt}{10pt}  [$10^5 M_\odot$] & [ckpc] \\ \hline \hline
    {\sc DaLya} fiducial & 38.4 & 7.2 & 1.0 \\ \hline
    {\sc DaLya} HR & 4.8 &  0.9 &  0.5 \\ \hline
    COLIBRE L050m6 & 24.2 &  18.4 & 1.8 \\ \hline
    COLIBRE L050m5 & 3.3 &  2.3 & 0.9 \\ \hline
    HELLO & 34.0 & 2.1 & 0.25 \\
    \hline 
  \end{tabular}\vspace{0.5cm}
\caption{Resolution of the simulations used in this work: dark matter mass, mean baryonic mass, and gravitational softening length for baryons.}
\label{table:sims}
\vspace{-0.5cm}
\end{table}

\begin{figure}
	\centering
	\includegraphics[scale=0.55]{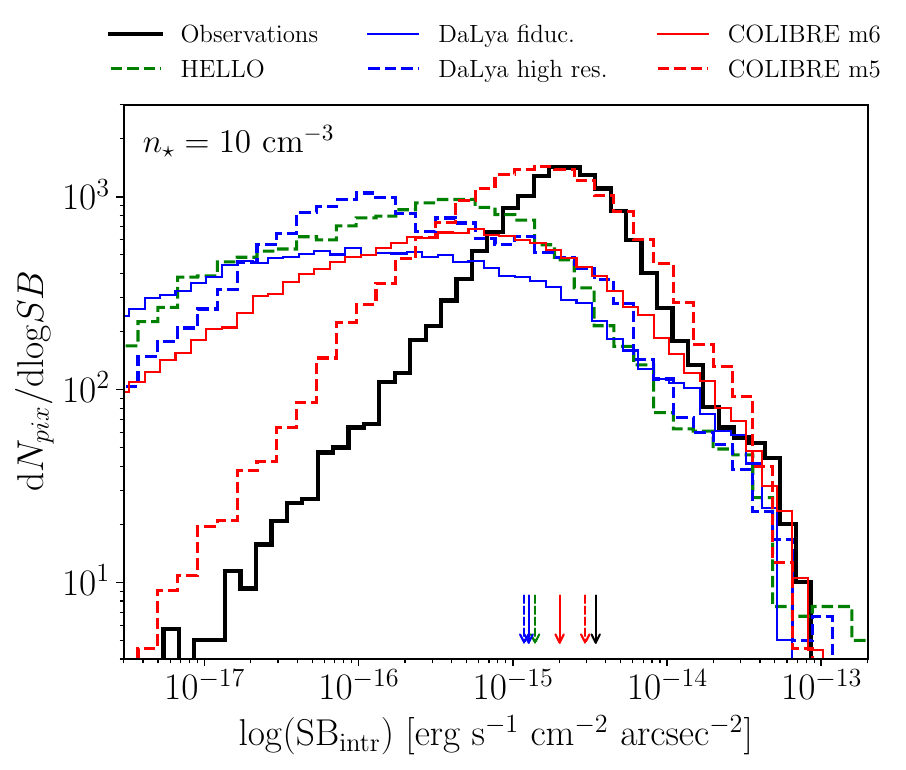}
    \caption{Comparison of the $\lya$ SB function with results from COLIBRE m6 (resolution similar to {\sc DaLya} fiducial) and from other simulations at 8 times higher resolution: {\sc DaLya} HR, COLIBRE m5, and HELLO. \dalya, COLIBRE and HELLO are presented in blue, red and green respectively. The linestyle (solid or dashed) indicates the resolution ($\sim 10^6$ and $\sim 10^5 M_\odot$ respectively). All simulations with the exception of COLIBRE m5 present PDF shapes which are strikingly different than the observed one, i.e. they predict a larger number of faint pixels than observed. We stress again that the simulated SB values should be considered as strict upper limits given our assumptions (see \refsec{MockOb}).}
    \label{fig:SB100200z3p5comp}
\end{figure}

Results for the $\lya$ SB function are presented in \reffig{SB100200z3p5comp}. All simulations with the exception of COLIBRE m5 (high resolution) display the same kind of distribution as \dalya fiducial and over-predict the number of faint pixels. Since \dalya HR, HELLO and COLIBRE m5 have all similar resolution, the striking difference between COLIBRE m5 and m6 (which can also be appreciated directly from the SB maps presented in \reffig{addmosaic}) cannot be simply ascribed to a resolution effect and hints at a nontrivial interplay between physics implementation and resolution on macroscopic observables. We return to this point in \refsec{discussion}.  

While our criterion for discarding ISM particles based on their density with a threshold at $n_\star = 10$ cm$^{-3}$ is well motivated in {\sc DaLya} and HELLO (since the same criterion is adopted in these simulations for star formation), it is less the case for COLIBRE. In \refapp{impactISM} we present the impact of removing the ISM on the $\lya$ SB function in COLIBRE m5, as well as different criteria to do so, showing that this only affects the very bright end of the histograms and does not affect our conclusions.

\begin{figure*}
	\centering
	\includegraphics[scale=0.5]{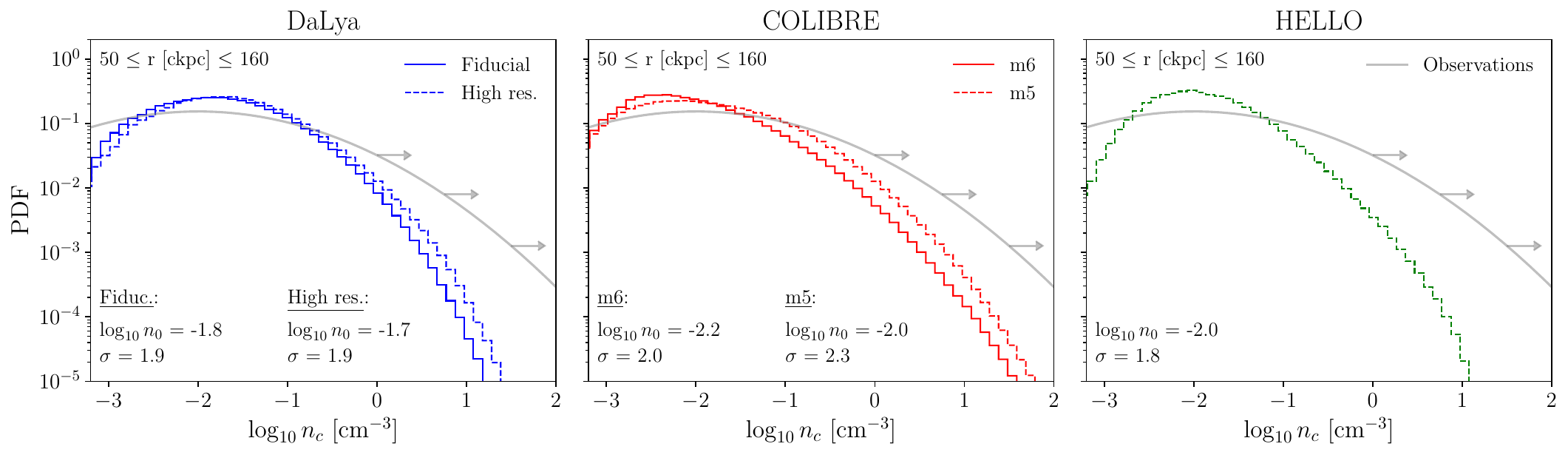}
    \caption{Same as \reffig{DensityPDF} but comparing to {\sc DaLya} HR (left panel), COLIBRE m6 and m5 (central panel), and HELLO (right panel), focusing on the 50--160 ckpc annulus only. Again, $n_0$ denotes the median density while $\sigma$ is the 1$\sigma$ equivalent width of $\ln n_c$. The corresponding mean density and clumping factor are listed in \reftab{meanclump}. The solid gray line shows the inferred PDF from the observations of the HeII/$\lya$ ratio from \citetalias{Travascio:inprep} in the same radial bin when fixing $\log_{10}n_0=-2$. The grey arrows illustrate that these are lower limit distributions.}
    \label{fig:DensityPDFHR}
\end{figure*}

Finally, we present the density PDFs of the cold gas (again without imposing any density threshold) for all simulations in \reffig{DensityPDFHR}, focusing on the 50--160 ckpc annulus only. The grey line in each panel again shows the PDF with the smallest $\sigma$ value compatible with observations of HeII/$\lya$ ratio when fixing the median density to $n_0=0.01$ cm$^{-3}$. As in \reffig{DensityPDF}, we quote the median density $n_0$ and 1$\sigma$ equivalent width of the distribution of $\ln n_c$. The histograms again display quasi-lognormal distributions skewed towards higher densities for all simulations (see \refapp{skewed}). As for \dalya fiducial, the simulated PDFs are narrower than that inferred from observations, which require a minimum $\sigma$ of 2.6 at $\log_{10}n_0=-2$, corresponding to a clumping factor of $\sim$860. In contrast, simulations have clumping factors between 16 and 102 as listed in \reftab{meanclump}, i.e. 8 -- 52 times lower. At fixed resolution, different simulations display different width of the PDF, with HELLO having the narrowest one and COLIBRE m5 the widest one (as could be expected from its different Ly$\alpha$ SB function). Moreover, the impact of resolution on $\sigma$ is very different in \dalya where it is unaffected, and COLIBRE where higher resolution results in a wider PDF. 

\begin{table}
\centering
  \begin{tabular}{ c | c | c | c | c  }
    \hline
    Sim.               & $\log_{10} n_0$ & $\sigma$ & \rule{0pt}{10pt} $\bar{n}$ [cm$^{-3}$] & $C_{\rm cold}$     \\ \hline \hline
    {\sc DaLya} fiduc. & -1.8            & 1.9      & 0.055  & 25.2  \\ \hline
    {\sc DaLya} HR     & -1.7            & 1.9      & 0.072  & 21.1  \\ \hline
    COLIBRE m6         & -2.2            & 2.0      & 0.037  & 29.0  \\ \hline
    COLIBRE m5         & -2.0            & 2.3      & 0.070  & 102.1 \\ \hline
    HELLO              & -2.0            & 1.8      & 0.034  & 16.5  \\
    \hline 
  \end{tabular}\vspace{0.5cm}
\caption{Parameters associated with the density PDFs presented in \reffig{DensityPDFHR}.}
\label{table:meanclump}
\vspace{-0.5cm}
\end{table}
By using CLOUDY photoionization models as discussed in \citetalias{Travascio:inprep}, we also calculated the expected HeII/$\lya$ ratio produced by the COLIBRE m5 PDF obtaining values between 0.12 and 0.23 considering a large possible range of distances and cloud sizes for the fiducial AGN spectrum with a EUV slope of -1.7. The predicted value of HeII/$\lya$ is at least twice as large as the measured HeII/$\lya$ ratio for the observed nebulae for which \citetalias{Travascio:inprep} obtain a value of 0.06$\pm$0.02. Considering the observational errorbars and a large range of model parameters, the predicted value is thus more than 2$\sigma$ away from current observational constraints. The predicted HeII/$\lya$ ratio would be equivalent to the one produced by a lognormal PDF with median density $n_0 =0.01$  and $2.5 \lesssim \sigma \lesssim 3.5$.

\section{Discussion} 
\label{sec:discussion}
We now discuss the implications of our results, taking advantage of our set of simulations using different subgrid models and resolution.

The impact of resolution on the small scale density of the CGM has been studied in numerous previous works (e.g. \citealt{Cantalupo2014, Hummels2019, Benett:2020, RameshNeslon2023, LiangRemming2020, Corlies2020, Augustin2021, Gronke2022}), showing that increasing resolution naturally increases the number of dense clouds. However, without additional physics at higher resolution the size function of cold clumps never converges \citep{RameshNeslon2023}. Things are further complicated by the fact that, in practice, the subgrid physics must be adjusted with resolution so that simulations remain calibrated against selected quantities as is done in COLIBRE \citep{Chaikin:2025}. This is not the case in \dalya for which the subgrid model was kept fixed, at the cost of slightly poorer agreement at higher resolution in the calibrated quantities. The results presented in \reffig{SB100200z3p5comp} show that the impact of resolution and its interplay with subgrid models on the macroscopic properties of the CGM such as the $\lya$ SB values is nontrivial. 
Indeed, the Ly$\alpha$ SB function is very similar for all simulations independent on resolution with the exception of COLIBRE m5. What is the origin of such large discrepancies? 

\begin{figure*}
\centering
\includegraphics[scale=0.5]{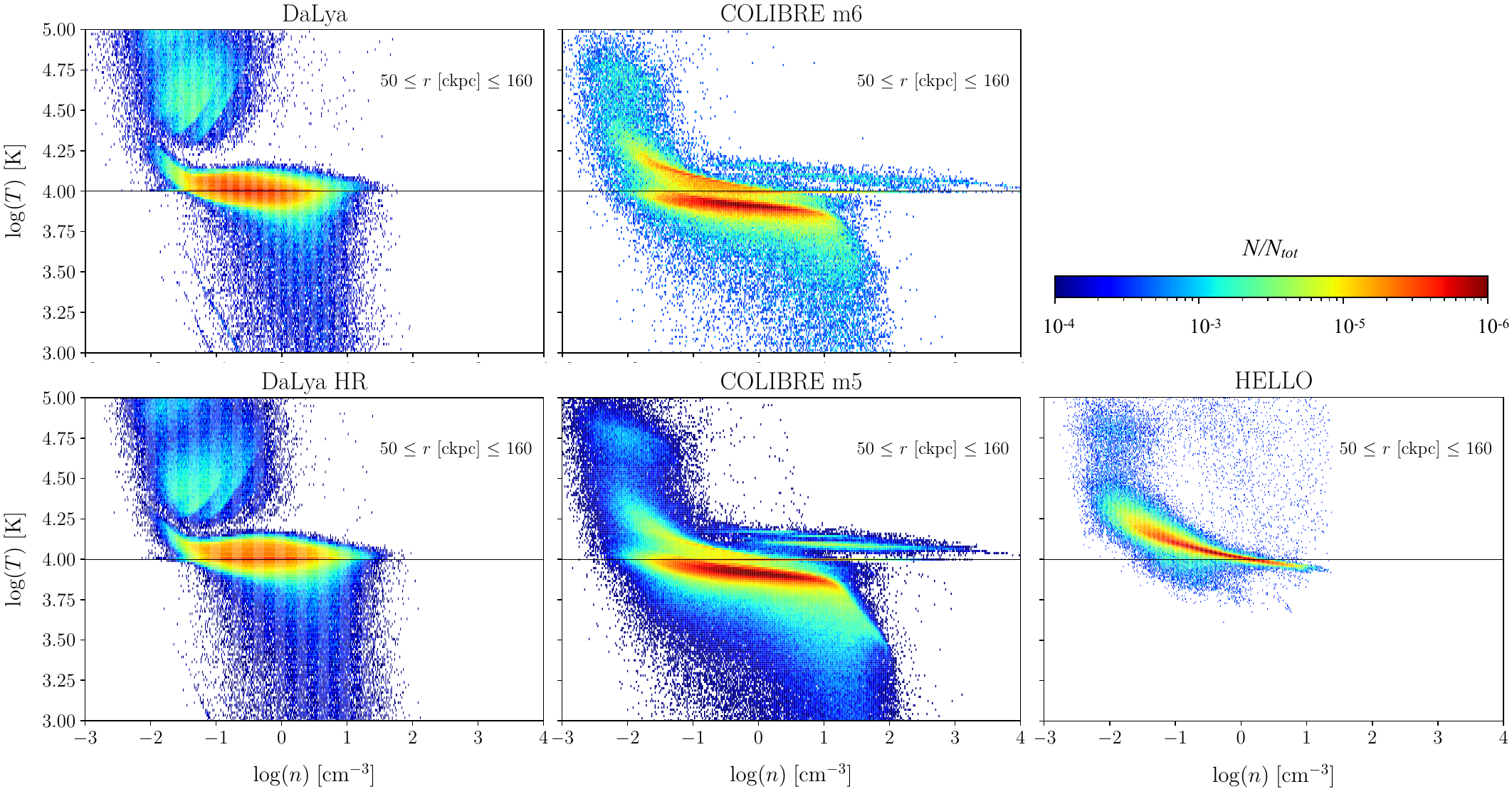}
\caption{The temperature -- density diagrams (mass weighted) at $z=3.5$ for particles within the 50 --160 kpc annulus in each simulations. The color-coding indicates the number density of particles in each $T$--$n$ bin. We focus on the temperature range $3 \leq \log T[K] \leq 5$. The black solid line indicates T=10$^4$K for reference.} 
\label{fig:trho}
\end{figure*}

While we found marginal evidence that energy injection from AGN feedback is reduced in m5 with respect to m6, the difference is less than 1$\sigma$ significant, and thus unlikely to solely explain the large difference observed in \reffig{SB100200z3p5comp}. Further insight can be gained from the mass weighted temperature -- density diagrams shown in \reffig{trho}, where we restrict to particles within 50 and 160 ckpc and in the temperature range 10$^3$ -- 10$^5$K. The diagrams display the usual cold CGM phase at $T \sim 10^{4}$K with densities roughly ranging from $10^{-2}$ to $10^2$ cm$^{-3}$. \dalya and COLIBRE further display low temperature, high density tails corresponding to the low temperature cooling in the ISM. COLIBRE also displays a high density tail slightly above $T=10^4$K corresponding to HII regions and low metallicity gas that has recombined from higher temperatures. 

In the density range between 0.01 and 10 cm$^{-3}$ \dalya displays a single peak temperature distribution roughly centered around 10$^4$K at both resolutions. HELLO also presents a single peak temperature distribution but the peak location strongly depends on the density. This difference most likely originates primarily on the different implementation of the UVB self-shielding and on different cooling functions. Indeed, these determine the equilibrium temperature of the CGM cold phase (before the quasars turns on) and thus its density. As explained before, \dalya adopts a simplified self-shielding following \cite{Rahmati:2013tr} with a self-shielding threshold at $\sim$0.1 cm$^{-3}$ for atomic gas. Since self-shielding is not implemented in HELLO, all particles are heated by the UVB even at densities around 1 cm$^{-3}$. COLIBRE includes self-shielding by dust and molecular gas in addition to atomic gas \citep{Ploeckinger:2025}, which is more dependent on the adopted resolution at the CGM densities probed in this work. As a consequence, COLIBRE m6 displays both phases with the temperature distribution at fixed density being double peaked reflecting the fact that only part of the gas with 0.01 $\lesssim$ $n$ $\lesssim$ 1 is self-shielded. We have further verified that the higher temperature density dependent peak is indeed constituted of almost fully ionized hydrogen while the lower temperature one is almost fully neutral. Interestingly enough, in COLIBRE m5 almost all particles with $n>0.1$ cm$^{-3}$ appear to be self-shielded and the temperature distribution at fixed $n$ is roughly single-peaked, resulting in a larger fraction of particles being cooler and denser in m5 than m6. Additional differences in the cooling prescription bring the median temperature of dense gas ($n \gtrsim 1$cm$^{-3}$) below 10$^4$K in COLIBRE. While a complete picture of the impact of resolution and different physical mechanisms on the $\lya$ SB values would require further investigation clearly out of the scope of this work, these considerations suggest that prescriptions for self-shielding and the adopted cooling functions might play an important role in the distribution and values of the $\lya$ SB in the CGM of quasars. 

Independently of these considerations, the comparison between the observed and simulated Ly$\alpha$ SB, which are upper limits due to the maximal fluorescence assumption and the lack of hydrodynamical response to the QSO photo-heating, demonstrate that densities of at least 10 cm$^{-3}$ should be present in the CGM of quasars at $z>3$. As this density is above the threshold used by several other cosmological simulations using an effective equation of state to define the ISM, our results hint that those simulations should either not impose an equation of state, or do so only at higher densities to properly model the CGM. Moreover, the results presented in \reffig{DensityPDFHR}, which are independent on the assumptions related to the comparison with the Ly$\alpha$ SB function, suggest that a broader density distribution than currently measured in simulations is required in order to match the observations, with the simulated clumping factor being 8 --52 times smaller than the one constrained from observed line ratios when fixing the same median density. Fixing the width of the distribution, on the other hand, leads to a median $\sim 10$ times higher and a mean $\sim 200$ higher than what is measured in simulations.

These results naturally raise questions regarding the formation of high-density cold substructures in the CGM. One possibility is that they could arise from instabilities at the interface of cold and hot gas, which can promote the condensation of cold gas \citep{Gronke2022}, its compression via compressive modes \citep{Mandelker2019}, and its fragmentation via shattering \citep{Yao:2025}. The scale at which this mechanism acts depends sensitively on the underlying microphysics: thermal conduction (e.g. \citealt{Ledos:2023}), viscosity (e.g. \citealt{Marin2025}), and magnetic fields (e.g. \citealt{Berlok:2019b, Ledos2024b, Das2024}). As we likely under-resolve the above instabilities and do not include physical mechanisms able to set a scale for the cold substructure, uncertainties remain in the exact shape of the density PDF. 

Assuming that at least the hot phase is resolved in our current simulations, we can obtain an estimate of the thermal pressure for each phase (as $P\sim n k_B T$). We use the diagrams in \reffig{trho} and compute the volume-weighted average temperature and density of the cold and hot phases. Our measurements reveal evidence for out of pressure equilibrium for all simulations, with $P_{\rm hot}/P_{\rm cold} \approx 8$ and 6 in {\sc Dalya} fiducial and HR, $\approx 10$ and 5 in COLIBRE m6 and m5, and $\approx 18$ in HELLO. The cold gas is thus \textit{on average} under-pressurized. While the exact impact of increasing the cold gas pressure is hard to predict due to several nonlinear effects, it could help bringing the simulated density PDF closer to constraints from observations. 

We note, however, that this apparent under-pressurization may also be, at least in part, a numerical artifact intrinsic to the way the SPH kernel interacts with nearby particles with a wide range of temperatures and densities. As is discussed in details in Appendix D of \cite{Oppenheimer:2018}, in the SPH scheme the calculation of cold clump densities is affected by neighboring lower density particles which have smoothing lengths overlapping the centers of cold clumps. This impacts the inferred values of the pressure of the individual particles, and, more importantly, this might also impact the simulated density distribution presented in figures \ref{fig:DensityPDF} and \ref{fig:DensityPDFHR}.

\subsection{Caveats}
\label{sec:caveats}

Our results depend on several assumptions that we discuss here. 

\textit{Maximal fluorescence:} The plausibility of this assumption depends on the cold gas in the CGM not being self-shielded from the QSO radiation, or, in other words, that the photoionization rate of the central quasar $\Gamma_i$ fulfills $\Gamma_i \gg n_c \times \alpha(T)$. As shown in \cite{Ledos:2026}, for a $10^{12}M_\odot$ halo, this condition is fulfilled in the whole CGM of bright QSOs as long as the density remains $\lesssim 60$cm$^{-3}$ at $r = R_{\rm vir}$ which, given the histograms shown in \reffig{DensityPDFHR}, is plausible for our case. Furthermore, \citet{ArrigoniBattaia2015} converted \cite{Borisova2016} MQN sample's $i$-band magnitudes to a photoionization rate of $\Gamma_i \simeq 10^{-7}(r/30\mathrm{kpc})^{-2} \mathrm{s}^{-1} \gg n_c \times \alpha(T)$. As the $i$-band magnitudes of the quasars contained in the observations sample that we use are comparable to those of MQN, their photoionization rates plausibly satisfy this assumption too.  

\textit{Ly$\alpha$ emission mechanisms:} We assume that the contribution to the emission due to collisional excitation is negligible compared to that of the recombination around bright QSOs as suggested by previous works \citep{Leibler2018,Langen2023}. \cite{Ledos:2026} verified this explicitly for various cases of clouds ionized by QSO radiation, and found recombination emission to be responsible for $> 90\%$ of the total $\lya$ emission in the optically thin limit. We stress in any case that a proper modeling of the contribution of collisional excitation is strongly dependent on both resolution and all the processes determining the ionization state and temperature of the gas, given the exponential dependence on temperature of the collisional excitation coefficient \citep{Cantalupo2008}. 

\textit{Quasar opening angle:} When generating the mock observations, we assume a 100\% opening angle of the ionizing torus due to the central quasar. While QSOs opening angle are still poorly constrained at $z >3$ \citep{Obreja2023}, a lower than $180^{\circ}$ opening angle would lead to fewer emitting regions and would increase the tension between simulations and observations, thus reinforcing our conclusions. 

\textit{Radiative transfer effects:} Even in the assumption of recombination dominating the emission, residual neutral hydrogen could result into local Ly$\alpha$ radiative transfer effects which would additionally ``smooth'' the mock observations, and change the shape of the simulated Ly$\alpha$ SB function. We explored this possibility by artificially smoothing the SB maps in {\sc DaLya} fiducial. However, in order to obtain simulated $\lya$ SB functions consistent with the observed one, the maps have to be smoothed with a gaussian kernel with a FWHM of 2.8 arcsec corresponding to $\sim$ 100 ckpc (22 pkpc), i.e, the full size of the annulus considered. Although we cannot completely rule out such large smoothing due to scattering, it seems highly implausible as it would cause the mock SB radial profiles to become much shallower than the observed ones. We note that Ly$\alpha$ scattering would not affect the comparison between the observed and simulated gas PDF as the former is constrained through the ratio with the non-resonant HeII line as discussed in \citetalias{Travascio:inprep}.

\textit{Constant photo-heating floor:} 
The equilibrium temperature between QSO heating and recombination cooling is unlikely to be constant throughout the entire CGM. Since our simulated SB values are upper limits and the recombination coefficient $\alpha(T)$ scales inversely with temperature, only an overestimation of the photo-heating floor would affect our results and conclusions. As \citet{PezzulliCantalupo2019} show through CLOUDY modeling, the equilibrium temperature depends on multiple variables, such as the gas metallicity, density, distance to the ionizing source and the shape of the ionizing spectrum. The equilibrium temperature scales inversely with metallicity for densities above $n \sim0.2$ cm$^{-3}$ and the temperature floor we impose corresponds to a metallicity of 0.1 Z$_{\odot}$. A lower photo-heating floor would correspond to even higher metallicites which, as shown in \cite{Popping:2022}, are predominantly found in the ISM rather than in the CGM at $z\geq3$. More recently, \cite{Ledos:2026} showed that the equilibrium temperature can be predicted from the ratio $F/(cn)$ (where $F$ is the ionizing photon flux), and found it to be $\sim 7\times 10^4$K for $n\sim 1$ cm$^{-3}$ in the optically thin limit, arguing that it is unlikely to depend strongly on metallicity. The equilibrium temperature also scales inversely with gas density. As our imposed floor is derived for a density of $n=1$ cm$^{-3}$, we are potentially overestimating the temperature of the denser gas ($n>10$ cm$^{-3}$), which could skew the SB histograms towards higher values if a lower equilibrium temperature was imposed on those very dense voxels. However, this is counterbalanced by the far more abundant cold gas exhibiting densities below $n=1$ cm$^{-3}$ which would increase the number of pixels at lower SB values. The net result would be to broaden the SB histograms even more without changing the covering fraction at high SB values, while bringing the simulated SB profiles down. 

\textit{No hydrodynamic response:}
The cold gas in the CGM also expands due to photo-heating from the central QSO. We do not model this hydrodynamic response, and use the gas densities as outputted by simulations to compute the $\lya$ SB and density histograms. However, as shown in \cite{Ledos:2026}, the density distribution of a cold cloud in the CGM of a QSO in the optically thin limit is shifted by at least an order of magnitude towards lower values after $\sim 5$Myr for an assumed cloud radius of 50 pc. This time is much shorter that the time needed by a photon to cross the entire nebulae at the speed of light, and gas dilation should thus be taken into account. As \cite{Ledos:2026} show, this dilation leads to the $\lya$ being only 5 -- 20\% of the maximum fluorescence value obtained from the gas initial density (see their figure 9). These considerations only reinforce our point that the simulated cold gas has to be on average denser or display a broader density distribution. 

\textit{Halo mass range:} Numerous studies find QSO host halos associated to $\lya$ nebulae having mass in the range $10^{12}$ -- $10^{12.5}$ $M_\odot$ at $2 \leq z \leq 5$ (see figure 7 of \citetalias{deBeer2023}). However, some studies favor higher masses ($M\sim 10^{13} M_\odot$) at $z\geq3$ \citep{Shen2007, Timlin2018}. As discussed in \citetalias{deBeer2023}, these estimates are obtained from QSO autocorrelation without taking nonlinear terms or shotnoise into account which most likely leads to an overestimate of the halo mass. We have verified that considering a broader or higher mass range is not sufficient to reconcile the simulated density PDFs shown in \reffig{DensityPDF} and \reffig{DensityPDFHR} with constraints from HeII/$\lya$ observations.

\section{Summary}
\label{sec:conclusion}
Making use of the constraints derived in \citetalias{deBeer2023} on the mass of dark matter halos hosting $\lya$ nebulae, this work focused on deriving constraints on the density distribution of the cold CGM by comparing the simulated and observed $\lya$ emission, as well as using constraints from the non-resonant line HeII-H$\alpha$. In addition to the newly developed \dalya simulations, we made use of HELLO and COLIBRE (m6 and m5) to span different resolution, hydrodynamic solvers, and implementations of astrophysical processes. These have been converted into mock observations similar to MUSE observations. 

Our main results can be summarized as follows:
\begin{itemize}
\item With the exception of COLIBRE m5 (high resolution), the simulated $\lya$ SB functions (defined as the average number of pixels per logarithmic SB bin) display strong differences with observations as they predict a larger number of low SB pixels. The results is that the $\lya$ covering fraction does not reach 100\% in simulations, differently than observations. 
\item Despite the very different shapes of the $\lya$ SB functions, the average $\lya$ SB values in \dalya is only within a factor of 3 from the observed value because of the dominant contribution of the brightest and rarer pixels. This suggests that \textit{comparing the circularly averaged SB values of simulations and observations cannot capture the full physics and properties of the CGM}.
\item The simulated cold gas density PDFs are close to skewed lognormal. To our knowledge, this is the first time that it was explicitly verified for the cold CGM. Constraints from observations which take advantage of the measured HeII/Ly$\alpha$ ratio and do not depend on the maximal fluorescence assumption are in tension with simulations. In particular, at fixed median density, constraints from observations require a clumping factor at least 8 times larger than the simulated one. 
\item Inspection of the $T-n$ diagrams hints that details in the self-shielding from the UVB and cooling might play an important role for sufficient cold dense gas to be present in simulations to reproduce the observed $\lya$ SB at a given resolution. 
\end{itemize}
Our results imply that \textit{the average density of the cold, $\lya$ emitting gas in the CGM, its clumpiness, or the volume filling factor (or all three) are too low in current simulations at $z\sim 3.5$}. We discussed the limitations of our physical assumptions, arguing that they do not strongly impact our conclusions. While we argued that small-scale turbulence might be one of the main physical process capable of modifying the simulated distribution to closer match observations, further work is necessary to investigate in details how it actually takes place in the CGM and how it impacts the clumping factor and the mean density. In a companion paper, \citetalias{Travascio:inprep} make use of the non-resonant emission lines H$\alpha$ and HeII to further constrain the density distribution of the cold CGM from an observational point of view, bringing valuable complementary insights to the results presented in this work. 

\section*{Acknowledgements}

TL thanks Alejandro Benitez-Llambay and Yuri Oku for useful discussions regarding the numerical setup of the {\sc DaLya} simulations. We acknowledge the CINECA award under the ISCRA initiative for the availability of high-performance computing resources and support. AVM and FB acknowledge support by Tamkeen under the NYU Abu Dhabi Research Institute grant CASS. This project was supported by the European Research Council (ERC) Consolidator Grant 864361 (CosmicWeb). HELLO simulations were carried out on the high-performance computing resources at New York University Abu Dhabi.

\bibliographystyle{aa} 
\bibliography{sources} 


\begin{appendix}
\section{Validation of the DaLya simulations}
\label{app:dalya}

\begin{figure}
\centering
\includegraphics[scale=0.5]{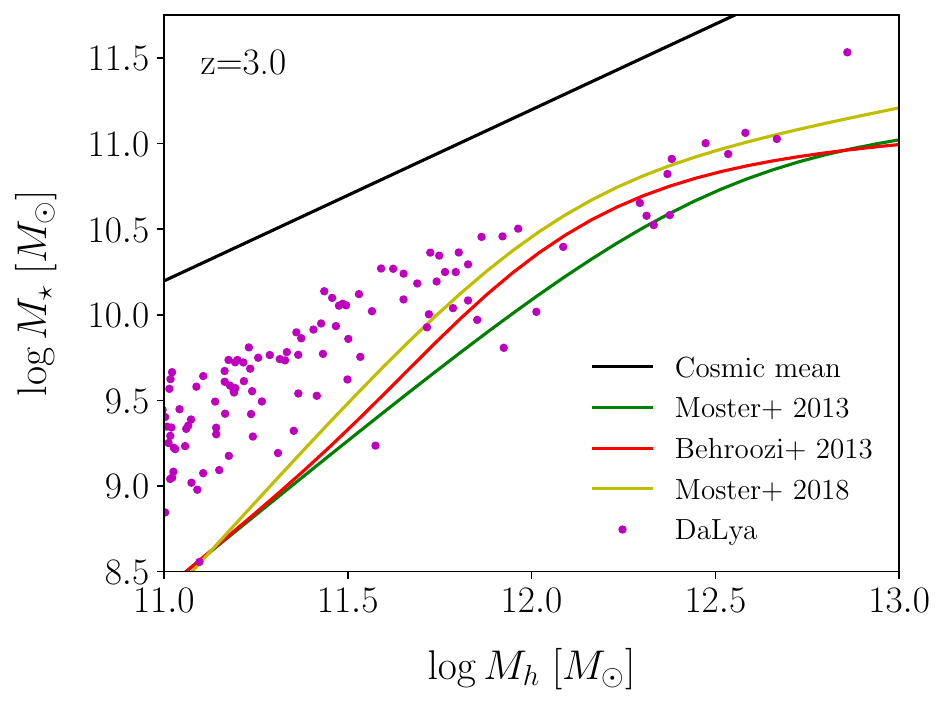}

\includegraphics[scale=0.5]{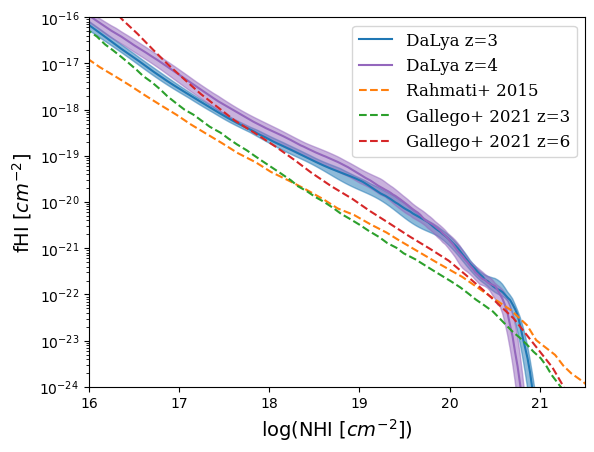}
\caption{\textbf{Top:} The central galaxy stellar mass to total halo mass relation at $z=3$ (magenta dots) and comparison with other relations from the literature. \textbf{Bottom:} The neutral hydrogen probability distribution function at $z = 4$ and $3$ (purple and blue curves respectively) compared to observational constraints at similar and higher redshifts.}
\label{fig:calibration}
\end{figure}

\begin{figure}
\centering
\includegraphics[scale=0.52]{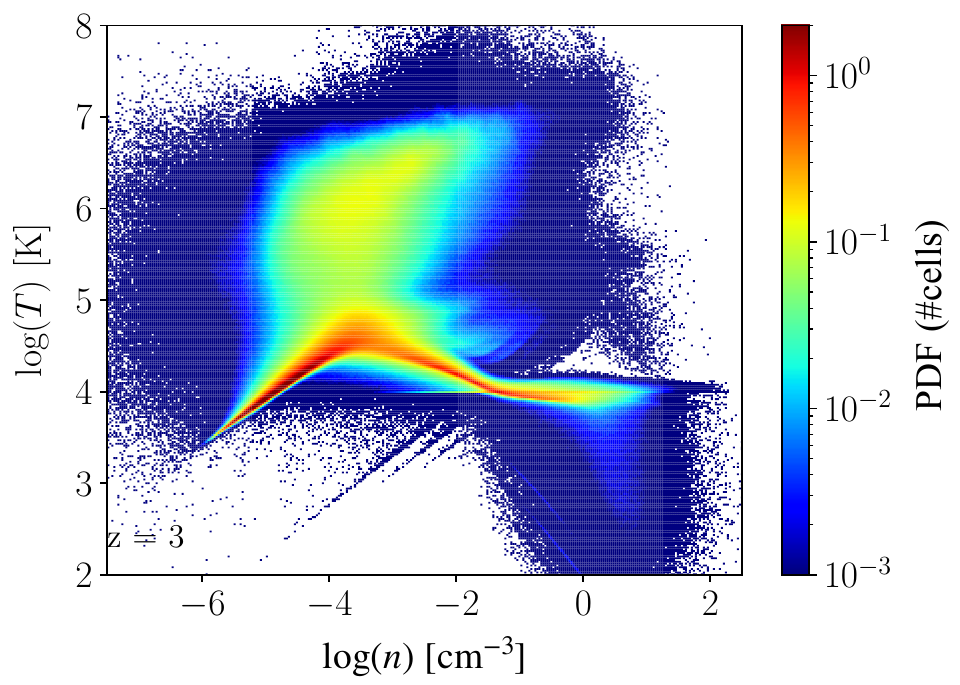}
\caption{\textbf{Top:} The temperature -- density diagram (mass weighted) in {\sc DaLya} fiducial at $z=3$ using all particles and combining both zoom-in regions.}
\label{fig:Trhofull}
\end{figure}

We present here the tests that we conducted to validate the {\sc DaLya} simulations using the fiducial resolution. We calibrate our model at $z=3$ against the central galaxy stellar mass to halo mass, and the neutral hydrogen column density probability distribution function. These are shown in \reffig{calibration}. In addition, we conduct a diagnostic of the temperature -- density diagram of the entire zoom-in region, shown in \reffig{Trhofull}. We stress again that the {\sc DaLya} simulation suite has not been designed to reproduce galaxy population and their properties on the smallest scales, but to study the CGM and the IGM. 

The top panel of \reffig{calibration} shows the stellar to halo mass relation at $z=3$ compared to known relations from the literature \citep{Moster:2013,Behroozi:2013,Moster:2018}. We also show the cosmic baryon fraction which represent the theoretical upper limit for which a galaxy with a baryon fraction equal to the cosmic one would transform all its gas into stars. The stellar to halo mass relation in {\sc DaLya} fiducial falls on top of other empirical relations for halos with mass $M\geq 10^{12} M_\odot$, crucially including the halo sample we use for this work. At lower masses, we observed a higher stellar mass than from other models. However, the measured stellar masses are at least one order of magnitude lower than the cosmic mean, ensuring that a large amount of baryons remain in the form of gas. 

The second quantity we calibrate our model against is the neutral hydrogen column density distribution function as defined in \cite{Tytler:1987}
\be
f_{HI} = \frac{{\rm d}N^2}{{\rm d}N_{HI}{\rm d}X},
\ee
with $N$ the number of atomic neutral hydrogen systems along the LOS, $N_{HI}$ is the neutral hydrogen column density and 
\be
{\rm d}X = \frac{H_0}{H(z)}(1+z)^2{\rm d}z.
\ee
We compare $f_{HI}$ at $z=4$ and $z=3$ to other relations from the literature \citep{Rahmati2015,Gallego:2021} in the lower panel of \reffig{calibration}. At low column density, the agreement between {\sc DaLya} and other work is excellent. We do observe a slight excess at column density $\sim 10^{19} -- 10^{20}$ cm$^{-2}$ which are self-shielded systems and might be attributed to the fact that we do not model radiation sources other than the homogeneous UVB. The sharp cutoff at very high $N_{HI}$ corresponds to our resolution limit.

Finally, we analyze the temperature -- density diagram of all high-resolution particles at $z=3$ in \reffig{Trhofull}. The diagram displays several regions corresponding to several phases of the gas standard in hydrodynamic simulations. At low densities ($\leq 10^{-4}$ cm$^{-3}$), most of the gas follows the imposed equation of state. At higher densities, a large part of the gas becomes efficiently cooled and lies on a metastable relation asymptoting to $10^4$K at densities $\geq 10^{-2}$ cm$^{-3}$, corresponding to the cold CGM as shown in the main text. At higher densities, cooling is so efficient that a small fraction of the gas can have temperature below $10^4$K, corresponding to the molecular phase. Finally, a non-negligible fraction of particle with densities between $10^{-5}$ and $10^{-1}$ cm$^{-3}$ is heated to roughly the virial temperature ($T\sim 10^6$K) and correspond to the hot CGM.

The calibration test and analysis conducted in this section show that the {\sc DaLya} simulation are in agreement with known relations for relevant quantities and display a well understood behavior, which validates them and their use for our purpose.

\section{Surface brightness maps}
\label{app:addmosaic}

\begin{figure*}
\centering
\includegraphics[scale=0.5]{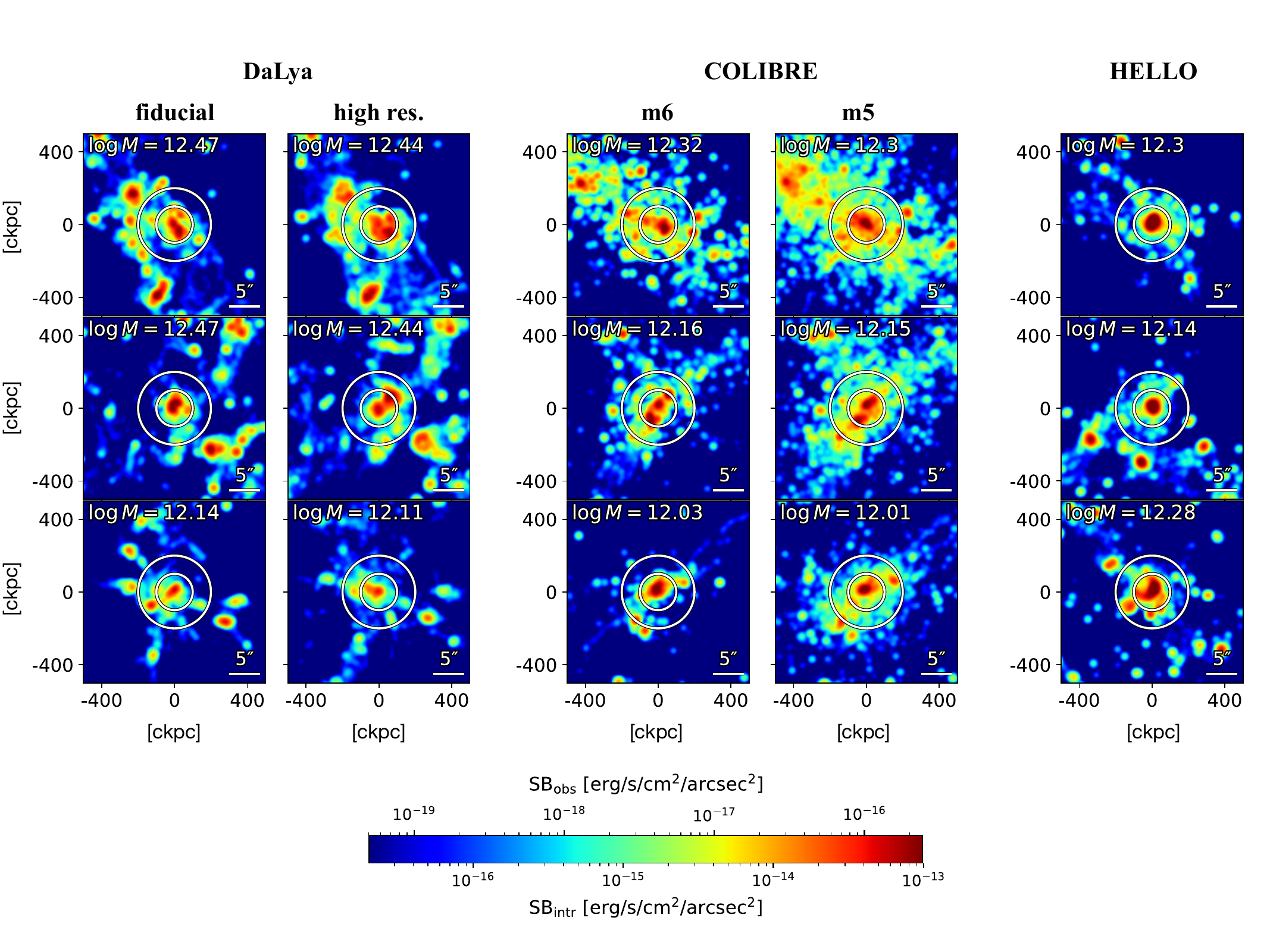}
\caption{$\lya$ surface brightness maps for each set of simulations. Each panel covers a 1x1 cMpc$^2$ field of view. Each column corresponds to a different simulation set, and each row is a different object. For \dalya and COLIBRE we show the same objects at two different resolutions to illustrate the impact of resolution. The two white concentric circles indicate the 100 -- 200 ckpc annulus used to study the $\lya$ SB function.}
\label{fig:addmosaic}
\end{figure*}

We provide here additional SB maps for a representative subset of 3 objects in each simulation set used in \refsec{simcomp}. These are shown in \reffig{addmosaic}. Each column corresponds to a different simulation set, and each row is a different object, covering the mass range used in this work. For \dalya and COLIBRE we show the same objects at two different resolutions to illustrate the impact of resolution. As for \reffig{mosaicSB}, the two white concentric circles indicate the 100-200 ckpc annulus used to study the $\lya$ SB function in the main text. As already pointed out in the main text, in \dalya fiducial, COLIBRE m6, and HELLO several pixels within the 100 -- 200 ckpc annulus are not detected as part of $\lya$ nebulae. Increasing the resolution in \dalya does not significantly affect the extent or shape of the nebulae. In contrast, $\lya$ nebulae identified in COLIBRE m5 are much more extended than those in m6, resulting in the 100 -- 200 ckpc annulus being entirely filled by the nebulae and a strong impact on the $lya$ SB function, as discussed in the main text. 

\section{Impact of removing the ISM in COLIBRE}
\label{app:impactISM}

\begin{figure}
\centering
\includegraphics[scale=0.55]{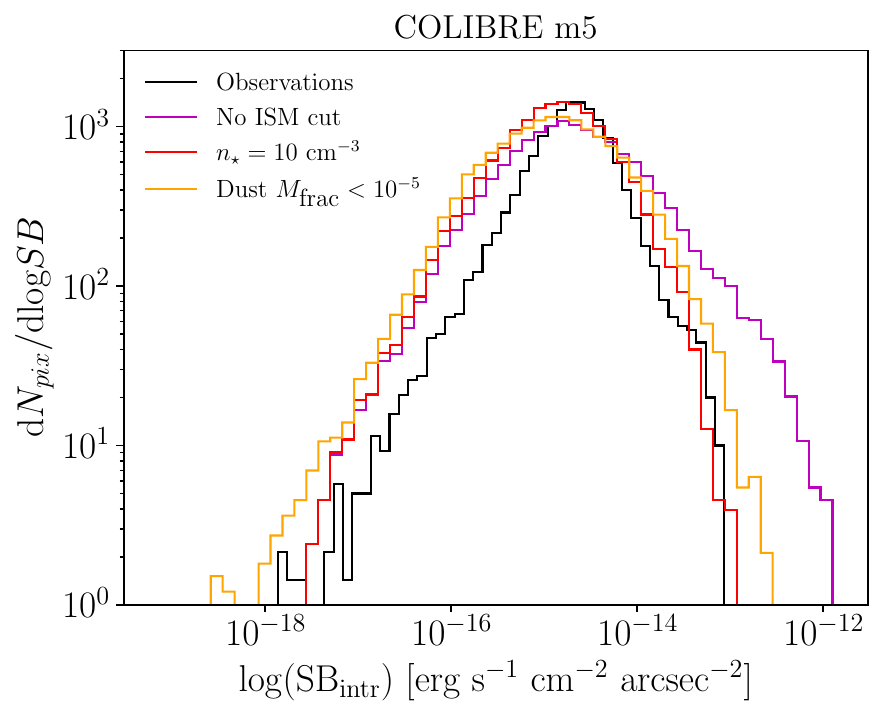}
\caption{Impact of the implementation of the ISM cut in COLIBRE m5 on the $\lya$ SB function. The red line shows result when removing the ISM based on the gas density, and corresponds to the dashed red line in \reffig{SB100200z3p5comp}. The other lines present results when no ISM cut is performed (purple), and when the ISM is removed based on the dust mass fraction (orange). The black line shows results from the observations by \citetalias{Travascio:inprep}.}
\label{fig:ISMLya}
\end{figure}

Our criterion for removing ISM cells based on their density is well motivated in \dalya and HELLO as these simulations use the same criterion to trigger star formation. This is not the case in COLIBRE, and we present here results for the $\lya$ SB functions in COLIBRE m5 when we do not remove the ISM. We also show results when we use another criterion based on the dust mass fraction, keeping only particles with a dust mass fraction smaller than $10^{-5}$ which roughly corresponds to the minimum value found in the ISM \citep{Popping:2022}. The results are shown in \reffig{ISMLya}. The red line shows result when removing the ISM based on the gas density, and corresponds to the dotted redline in \reffig{SB100200z3p5comp}. The other lines present results when no ISM cut is performed (purple), and when the ISM is removed based on the dust mass fraction (orange). This last criterion is physically motivated as dusty regions can be expected to have suppressed $\lya$ emission due to dust attenuation. As expected, removing the ISM based on the gas density only affects the high SB end of the distribution. Results when removing the ISM based on dust are almost identical to those obtain with a density cut, indicating that high density particles indeed correspond to high dust fraction ones, and confirming the robustness of our results. 

In analogy with what was done in \reftab{meanclump} in \refsec{simcomp}, we also quote the parameters of the density PDFs associated to these diverse ISM cuts in \reftab{meanclumpISM}. The case without ISM cut corresponds to what was presented in the main text. As expected, removing the ISM results in a slightly lower mean density and strongly diminishes the clumping factor (and $\sigma$). In contrast it almost doesn't affect the median. 
\begin{table}
\centering
  \begin{tabular}{ c | c | c | c | c  }
    \hline
    ISM cut criterion               & $\log_{10} n_0$ & $\sigma$ & \rule{0pt}{10pt} $\bar{n}$ [cm$^{-3}$] & $C_{\rm cold}$     \\ \hline \hline
    No ISM cut                      & -2.0            & 2.3      & 0.0696  & 102.1  \\ \hline
    $n_\star = 10$ cm$^{-3}$        & -2.0            & 1.9      & 0.0655  & 18.5  \\ \hline
    Dust $M_{\rm frac} = 10^{-5}$   & -2.0            & 2.1      & 0.0657  & 38.5  \\ \hline
  \end{tabular}\vspace{0.5cm}
\caption{Parameters associated with the density PDFs corresponding to different ways of removing the ISM in COLIBRE m5.}
\label{table:meanclumpISM}
\vspace{-0.5cm}
\end{table}

\section{Skewed lognormal distribution}
\label{app:skewed}

As shown in \refsecs{denDistCGM}{simcomp}, the PDFs of the cold gas density display quasi-lognormal behavior. We explore the shape of the PDFs in more details here. We attempt to fit the PDFs of the natural logarithm of the cold gas density $n_c$ with a skewed lognormal distribution following \cite{Azzalini:1985}. Compared to the standard lognormal one, it assumes an additional parameter $\alpha$ describing the skewness of the distribution
\begin{figure}
\centering
\includegraphics[scale=0.55]{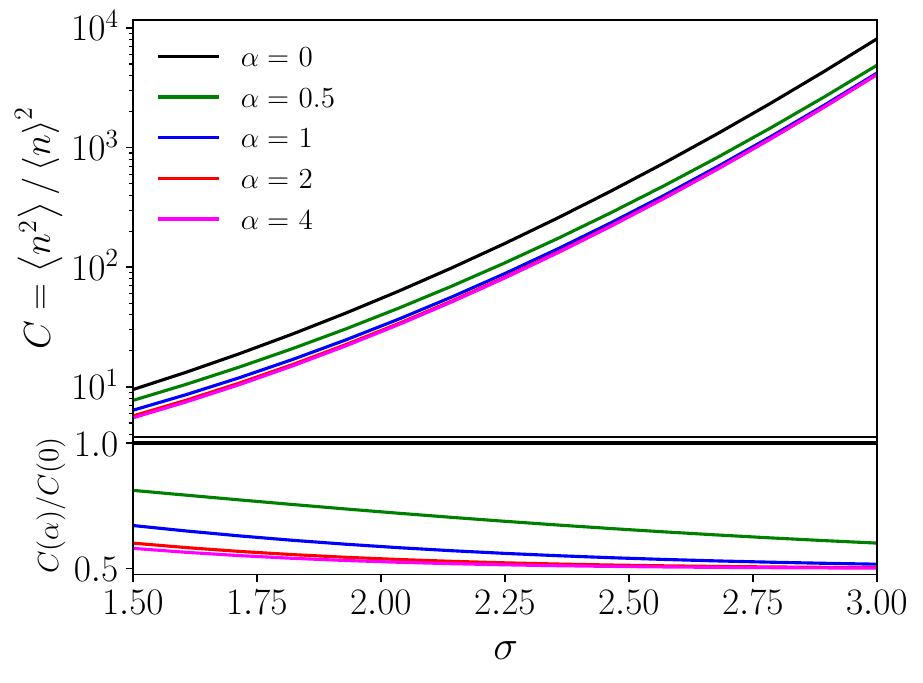}
\caption{Clumping factor for the lognormal (black) and skewed lognormal (colors) PDFs. We show the relations for several values of $\alpha$ compatible with our results. The bottom panel shows the ratio of C from the skewed to the lognormal PDF. The ratio is almost constant slightly under 1 for all $\sigma$ values considered here.}
\label{fig:clumping}
\end{figure}

\begin{figure}
\centering
\includegraphics[scale=0.6]{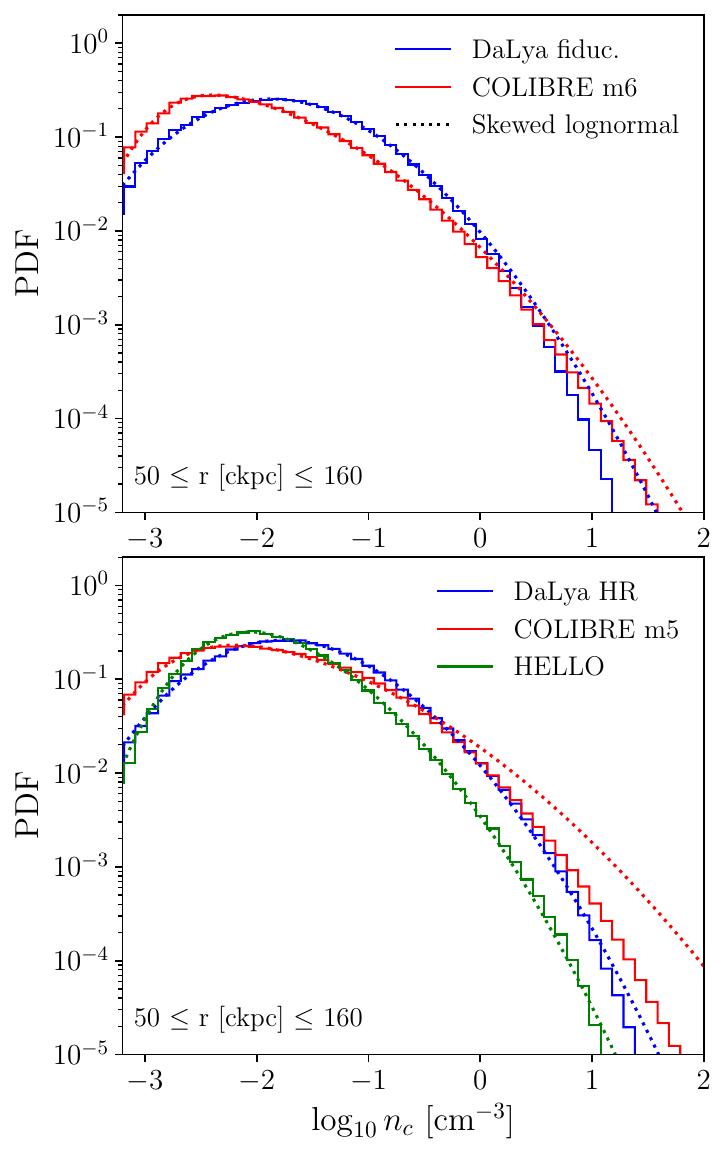}
\caption{PDF of the density distribution for the 5 sets of simulations considered in this work (each panel regroups simulations at roughly the same resolution), and associated best skewed-lognormal fits (dotted lines).}
\label{fig:PDFwFit}
\end{figure}

\begin{align}
{\rm PDF}(\ln n, \mu, \sigma, \alpha)\, \rm{d} \ln n & = \nonumber \\  = \frac{1}{\sqrt{2\pi\sigma^2}}e^{-\frac{(\ln n -\mu)^2}{2\sigma^2}} & \cdot \left[1+\rm{erf}\left(\frac{\alpha(\ln n - \mu)}{\sqrt{2\sigma^2}}\right)\right]\rm{d}\ln n,
\label{eq:skewedgaussian}
\end{align}
where $\mu$ is the mean of $\ln n$, $\sigma$ the width of the distribution and $\alpha$ the skewness, while the error function is defined as
\be
\rm{erf}(x) = \frac{2}{\sqrt{\pi}} \int_0^x e^{-t^2}\rm{d}t.
\ee

The first and second moments of the distribution can be computed analytically as
\begin{align}
\left\langle n \right \rangle = & \; e^{\mu +\sigma^2/2}\left[1 + {\rm erf}{\frac{\alpha\sigma}{\sqrt{2(1+\alpha^2)}}}\right], \nonumber \\
\left\langle n^2 \right \rangle = & \; e^{2\mu +2\sigma^2}\left[1 + {\rm erf}{\frac{\sqrt{2}\alpha\sigma}{\sqrt{1+\alpha^2}}}\right],
\end{align}
such that the clumping factor is
\be
C = e^{\sigma^2} \cdot \frac{1+{\rm erf(2u)}}{(1+{\rm erf}(u))^2}
\label{eq:clumpingSKLN}
\ee
where we defined $u = \alpha\sigma/\sqrt{2(1+\alpha^2)}$. The clumping factor as a function of $\sigma$ is shown in \reffig{clumping} for various values of $\alpha$ including 0 (lognormal case). The difference due to the skewness is relatively small and almost $\sigma$-independent, such that $C$ varies over several orders of magnitude when $\sigma$ doubles for all values of $\alpha$.

The fits are shown as dotted lines in \reffig{PDFwFit} for the 50-160 ckpc annulus. Each panel shows results at similar resolution. While the skewed lognormal fit performs extremely well for densities lower than 1 atoms per cm$^3$, it fails at high densities in several case, particularly \dalya fiducial and COLIBRE m5 where the simulated PDFs decrease faster that the analytical form. For this reason we refrain from using these fits to obtain the median density and width $\sigma$, and derive them directly from the measured density PDFs.


\label{lastpage}
\end{appendix}

\end{document}